\documentclass[11pt,a4paper]{scrartcl}

\usepackage{CLICdp}
\usepackage{float}
\usepackage{overpic}

\usepackage{CLICdp_definitions}

\title{Search for exotic long-lived particles at CLIC}

\clicdppub{2022}{002}  

\date{\today}

\addauthor{Marcin Kucharczyk}{\institute{1}\hcomma\editor{mateusz.goncerz@cern.ch,marcin.kucharczyk@cern.ch}}
\addauthor{Mateusz Goncerz}{\institute{1}\hcomma\editor{mateusz.goncerz@cern.ch}}

\addinstitute{1}{The Henryk Niewodniczanski Institute of Nuclear Physics, Polish Academy of Sciences, Krakow, Poland}

\abstract{A study of the sensitivity of the CLIC\_ILD detector model for massive long-lived particles produced in the decay of the Higgs boson is presented, using a data sample of $\textrm{e}^+\textrm{e}^-$ collisions at $\sqrt{\textrm{s}}=350$ GeV and $\sqrt{\textrm{s}}=3$ TeV, corresponding to an integrated luminosity of 1~ab$^{-1}$ and 3~ab$^{-1}$, respectively. The sensitivity range covers long-lived particle lifetimes from 1 to 300 ps, masses between 25 and 50 GeV, and a parent Higgs mass of 126 GeV. Sensitivities to the production cross-section as a function of the long-lived particle mass and lifetime are determined.}

\titlecomment{This work was carried out in the framework of the CLICdp Collaboration} 

\graphicspath{ {./logos/}{./figures/} }

\begin{document}

\titlepage


\section{Introduction}
Despite the great success of Standard Model of particle physics (SM) in describing physics processes at very short distances, the SM has a number of open issues, such as the hierarchy problem, or the absence of candidates for dark matter. Among many theoretical descriptions of new phenomena beyond the Standard Model there is a class of models predicting the existence of new massive Long-Lived Particles (LLP) with a measurable flight distance. One class of such models, the so-called Hidden Valley~\cite{hidValley1,hidValley2}, is a consequence of string-theory and predicts the existence of exotic particles arising from the introduction of an additional gauge sector. These additional gauge sectors can only be excited in high energy collisions. This energy is needed to overcome the energy barrier separating the Standard Model sector from the hidden valley sector, where the communicator can be, for example, a Z' boson or the Higgs boson. The Hidden Valley particles are predicted in some of these models to have non-zero lifetime and to decay into $b\bar{b}$, having unobservable partners that could serve as dark matter objects. They produce displaced vertices (DV) which can be efficiently reconstructed by the tracking system of the CLIC detector~\cite{clicILD,CLICDet}, increasing the sensitivity for massive long-lived particles. In the present paper the analysis of the Standard Model Higgs boson decaying into two Hidden Valley particles is described, $H \to \pi_\textrm{v}^\textrm{0} \pi_\textrm{v}^\textrm{0} \to b\bar{b} b\bar{b}$, providing four $b$-jets in the final state. At the $\sqrt{s}=350$~GeV the Higgs boson is dominantly produced in the Higgsstrahlung process ($\textrm{e}^+\textrm{e}^- \to ZH$), while at $\sqrt{s}=3$~TeV the dominant production mechanism is  $WW$-fusion. Both are simulated with the CLIC\_ILD detector model~\cite{CLICD}, assuming an integrated luminosity of 1~ab$^{-1}$ at $\sqrt{s}=350$~GeV and 3~ab$^{-1}$ at $\sqrt{s}=3$~TeV. Similar searches of the SM Higgs boson decaying into two LLP's providing two $b\bar{b}$ di-jets in the final state have already been reported by the D0~\cite{hvD0}, CDF~\cite{hvCDF}, ATLAS~\cite{hvATLAS}, CMS~\cite{hvCMS} and LHCb~\cite{hvLHCb} experiments.

\section{CLIC detector concepts}
The Compact Linear Collider (CLIC~\cite{clicILD}) project has developed accelerating gradients to extend $\textrm{e}^+\textrm{e}^-$ collisions into the multi-TeV regime. Several  detector frameworks have been used for CLIC physics studies, for example the CLIC\_SiD and CLIC\_ILD detector concepts, that were derived from the ILD and SiD detectors for ILC~\cite{clicSiD1,clicSiDILD,clicILD2}. As the most challenging environment for the detectors is related to the 3~TeV CLIC operation mode, detector systems are designed to meet the requirements associated with high beam-induced background levels at 3~TeV. Therefore, they are suitable at all energy stages, although the inner tracking detectors and vertex detector may be implemented with lower inner radius in case of $\sqrt{s}=350$~GeV, where the backgrounds are significantly lower. The CLIC detectors have been designed to fulfill several requirements of the experimental physics program, such as an excellent track-momentum resolution, precise impact parameter resolution, good jet-energy resolution for light-quark jet energies, and detector coverage for electrons extending to very low angles with respect to the beam axis, to maximize background rejection for $WW$-fusion events. This makes the CLIC detector an excellent tool to examine especially the Higgs boson, but also the existence of exotic long-lived particles. The present studies are based on the CLIC\_ILD detector concept in the first stage at $\sqrt{s}=350$~GeV and the ultimate stage at $\sqrt{s}=3$~TeV. At the first stage an integrated luminosity of 1~ab$^{-1}$ is assumed to be collected, while at the second stage 3~ab$^{-1}$ is assumed.

\section{Event generation and detector simulation}
The signal samples as well as the main physics backgrounds are generated using the Whizard~1.95~\cite{Whizard} program. The process of fragmentation and hadronization is simulated using Pythia~6.4~\cite{Pythia}, configured to produce Hidden Valley processes. The mass of the Higgs boson is taken to be 126~GeV/c$^{2}$, and the events from the different Higgs production processes are simulated separately. The interaction of the generated particles with the CLIC\_ILD detector and its response are implemented using the Geant4~\cite{Geant4} simulation package and the MOKKA~\cite{Mokka} detector description toolkit. Finally, the MARLIN software package~\cite{Marlin} is used for event reconstruction, with the track reconstruction as in~\cite{clicTrack}. The signal samples of $\textrm{e}^+\textrm{e}^-$ collisions at $\sqrt{s}=350$~GeV and $\sqrt{s}=3$~TeV, namely $\textrm{e}^+\textrm{e}^- \to ZH (H \to \pi_\textrm{v}^\textrm{0} \pi_\textrm{v}^\textrm{0})$ and $\textrm{e}^+\textrm{e}^- \to H \nu_{e} \bar{\nu_{e}} (H \to  \pi_\textrm{v}^\textrm{0} \pi_\textrm{v}^\textrm{0})$ respectively, with $\pi_\textrm{v}^\textrm{0}$ lifetimes from 1 to 300 ps, masses between 25 and 50 GeV/c$^{2}$, and a parent Higgs mass of 126 GeV/c$^{2}$ were generated. Background samples of $q \bar{q}$, $q \bar{q}\nu\bar{\nu}$, $q \bar{q} q \bar{q}$, $q \bar{q} q \bar{q} \nu\bar{\nu}$ were generated, with additional samples of $t \bar{t}$ and $WWZ$ for $\sqrt{s}=350$~GeV. The signal and background samples produced are summarized in Table~\ref{tab:samples}, where for every signal sample a cross section of 0.93~pb ($\sqrt{s}=350$~GeV) and 0.42~pb ($\sqrt{s}=3$~TeV) with $\textrm{BR}(\pi_\textrm{v}^0 \to b\bar{b}) = 100\%$ were assumed. In each case beam induced $\gamma \gamma \to hadrons$ interactions are overlaid for each event (see Ref.~\cite{hgg} for details) . The effect of beamstrahlung and initial state radiation results in a tail in the distribution of the effective centre-of-mass energy.

\begin{table}[h]
    \small
    \centering
    \begin{tabular}{c c|c c c|c c c}
    \multicolumn{2}{c}{} & \multicolumn{3}{c}{$\sqrt{s} = 350$~$\textrm{GeV}$} &
    \multicolumn{3}{c}{$\sqrt{s} = 3$~$\textrm{TeV}$}\\
    \bottomrule
    $m_{\pi_\textrm{v}^\textrm{0}}$[GeV] & $\tau_{\pi_\textrm{v}^\textrm{0}}$[ps] & $\sigma$[pb] & sample size & Eff.$^{presel.}$ [\%] & $\sigma$[pb] & sample size & Eff.$^{presel.}$ [\%] \\
    \midrule
    25 & 1 & 0.93 & $\sim 240 \textrm{K}$ & 78 & 0.42 & $\sim 200 \textrm{K}$ & 68\\
    25 & 10 & 0.93 & $\sim 240 \textrm{K}$ & 94 & 0.42 & $\sim 200 \textrm{K}$ & 86\\
    25 & 100 & 0.93 & $\sim 240 \textrm{K}$ & 99 & 0.42 & $\sim 200 \textrm{K}$ & 93\\
    25 & 300 & 0.93 & $\sim 240 \textrm{K}$ & 97 & 0.42 & $\sim 200 \textrm{K}$ & 80\\
    \midrule
    35 & 1 & 0.93 & $\sim 240 \textrm{K}$ & 76 & 0.42 & $\sim 200 \textrm{K}$ & 70\\
    35 & 10 & 0.93 & $\sim 240 \textrm{K}$ & 93 & 0.42 & $\sim 200 \textrm{K}$ & 86\\
    35 & 100 & 0.93 & $\sim 240 \textrm{K}$ & 99 & 0.42 & $\sim 200 \textrm{K}$ & 94\\
    35 & 300 & 0.93 & $\sim 240 \textrm{K}$ & 98 & 0.42 & $\sim 200 \textrm{K}$ & 82\\    
    \midrule
    50 & 1 & 0.93 & $\sim 240 \textrm{K}$ & 72 & 0.42 & $\sim 200 \textrm{K}$ & 72\\
    50 & 10 & 0.93 & $\sim 240 \textrm{K}$ & 89 & 0.42 & $\sim 200 \textrm{K}$ & 89\\
    50 & 100 & 0.93 & $\sim 240 \textrm{K}$ & 99 & 0.42 & $\sim 200 \textrm{K}$ & 90\\
    50 & 300 & 0.93 & $\sim 240 \textrm{K}$ & 99 & 0.42 & $\sim 200 \textrm{K}$ & 86\\
    \midrule
    \multicolumn{2}{c|}{q$\bar{\textrm{q}}$} & 24.41 & $\sim 2 \textrm{M}$ & 12 & 2.95 & $\sim 200 \textrm{K}$ & 6\\
    \multicolumn{2}{c|}{q$\bar{\textrm{q}}\nu\bar{\nu}$} & 0.32 & $\sim 306 \textrm{K}$ & 12 & 1.32 & $\sim 200 \textrm{K}$ & 8\\
    \multicolumn{2}{c|}{q$\bar{\textrm{q}}$q$\bar{\textrm{q}}$} & 5.85 & $\sim 1.44 \textrm{M}$ & 8 & 0.55 & $\sim 750 \textrm{K}$ & 9\\
    \multicolumn{2}{c|}{q$\bar{\textrm{q}}$q$\bar{\textrm{q}}\nu\bar{\nu}$} & \multicolumn{3}{c|}{-} & 0.07 & $\sim 300 \textrm{K}$ & 11\\
    \multicolumn{2}{c|}{t$\bar{\textrm{t}}$} & 0.45 & $\sim 241 \textrm{K}$ & 12 &\multicolumn{3}{c}{-} \\
    \multicolumn{2}{c|}{WWZ} & 0.01 & $\sim 40 \textrm{K}$ & 14 &\multicolumn{3}{c}{-} \\
    \bottomrule
    \end{tabular}
    \caption{Signal and background Monte Carlo samples. Eff.$^{presel.}$ is the preselection efficiency denoting the efficiency of the events with at least two displaced vertices reconstructed and a $b$-tag probability of $>$~0.95 for all reconstructed jets (see Sect.~4).}
    \label{tab:samples}
\end{table}

\section{Analysis procedure}
\begin{figure}[ht]
\begin{center}
\begin{overpic}[width=0.405\linewidth]{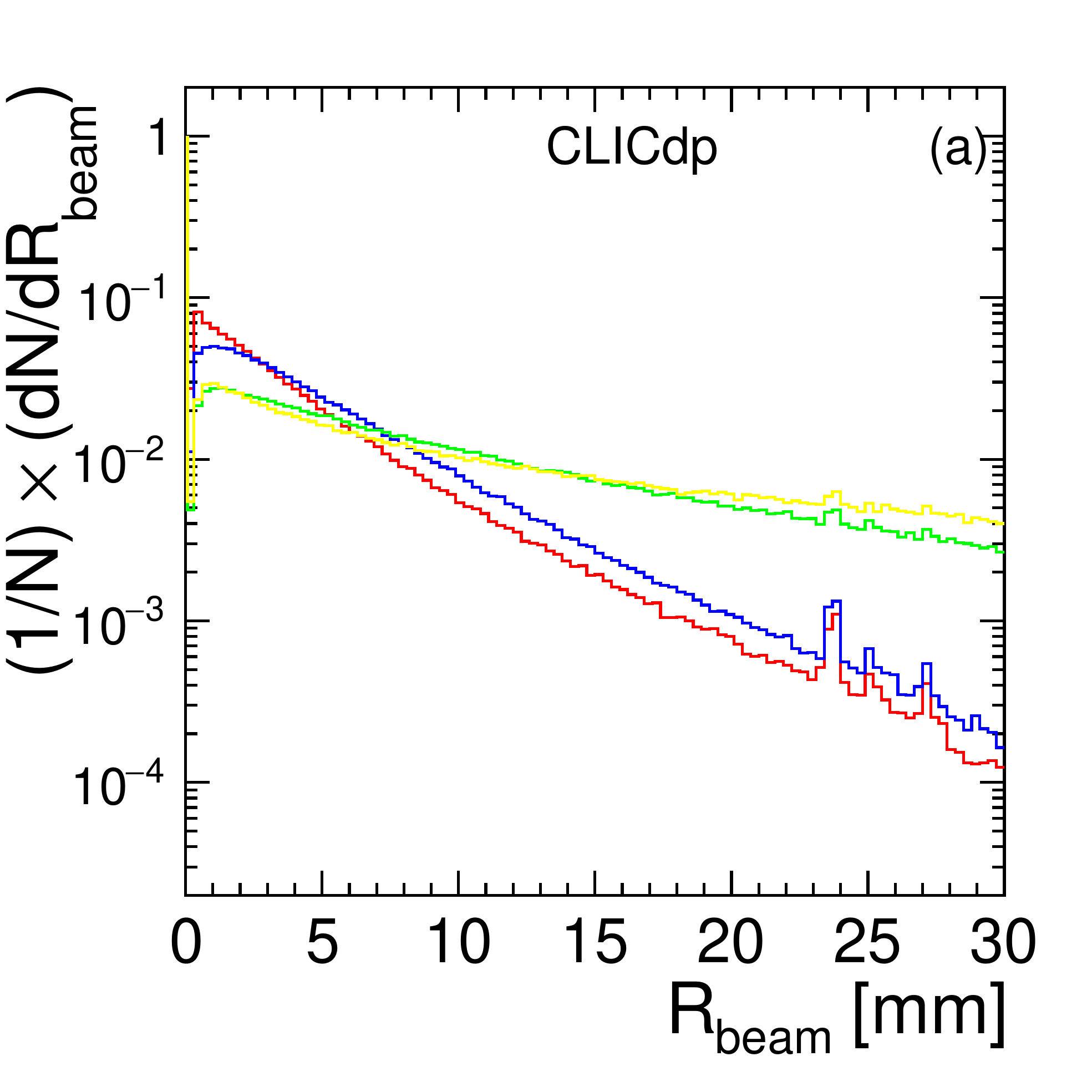}
\put(45,80){\scriptsize$\sqrt{s} = 350$ GeV}%
\end{overpic}
\begin{overpic}[width=0.42\linewidth]{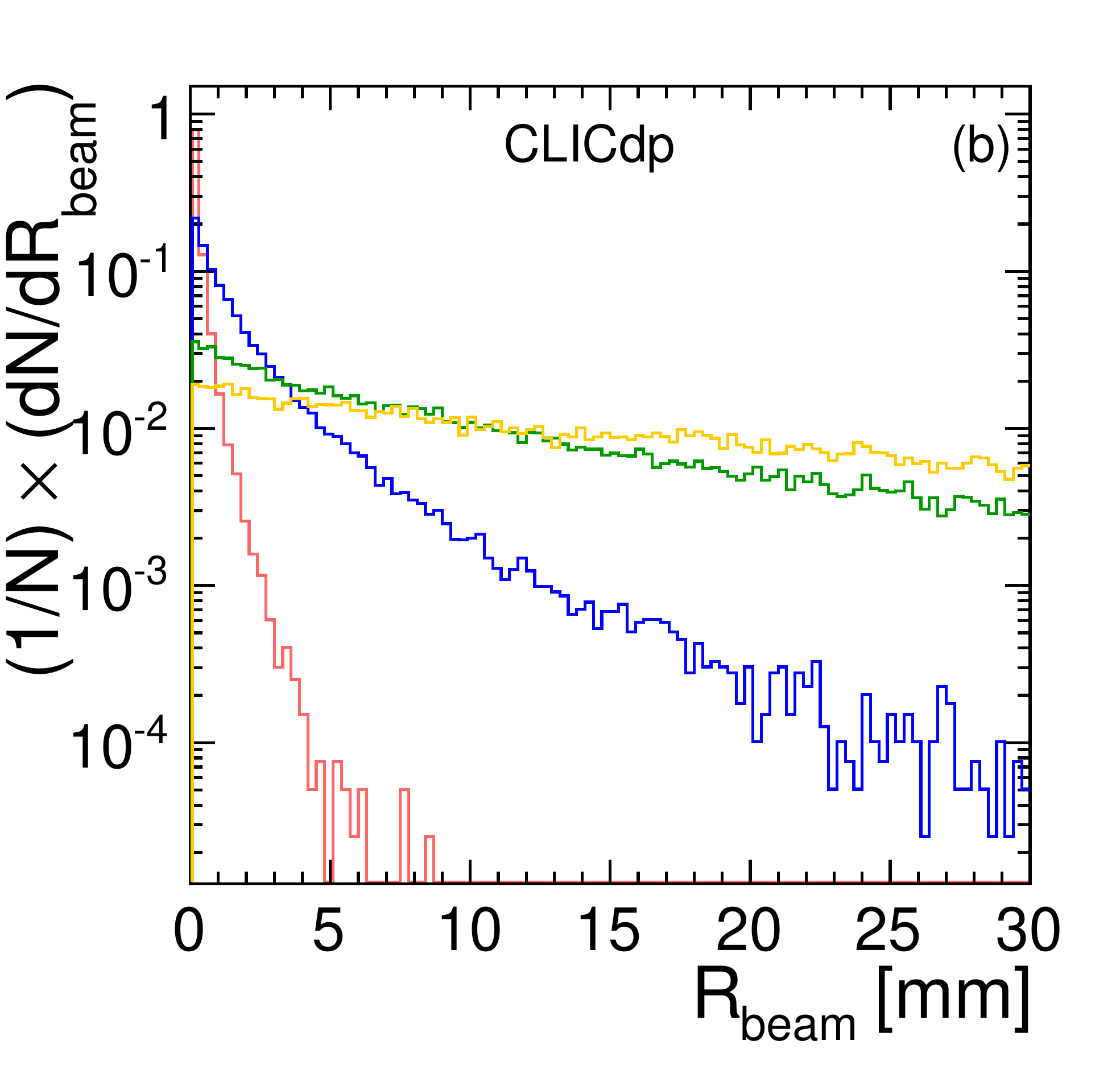}
\put(42,77){\scriptsize$\sqrt{s} = 3$ TeV}%
\end{overpic}
\end{center}
\caption{Radial distance of the generated $\pi^0_v$ to the beam axis for $\pi^0_v$'s generated with a mass of 50 GeV/c$^{2}$ and with four different lifetimes: 1~ps (red), 10~ps (blue), 100~ps (green) and 300~ps (yellow)~\cite{CLICHidValley} for (a) $\sqrt{s}=350$~GeV and (b) $\sqrt{s}=3$~TeV.}
\label{fig:genHVr}
\end{figure}

\begin{figure}[ht]
\begin{center}
\begin{overpic}[width=0.37\linewidth]{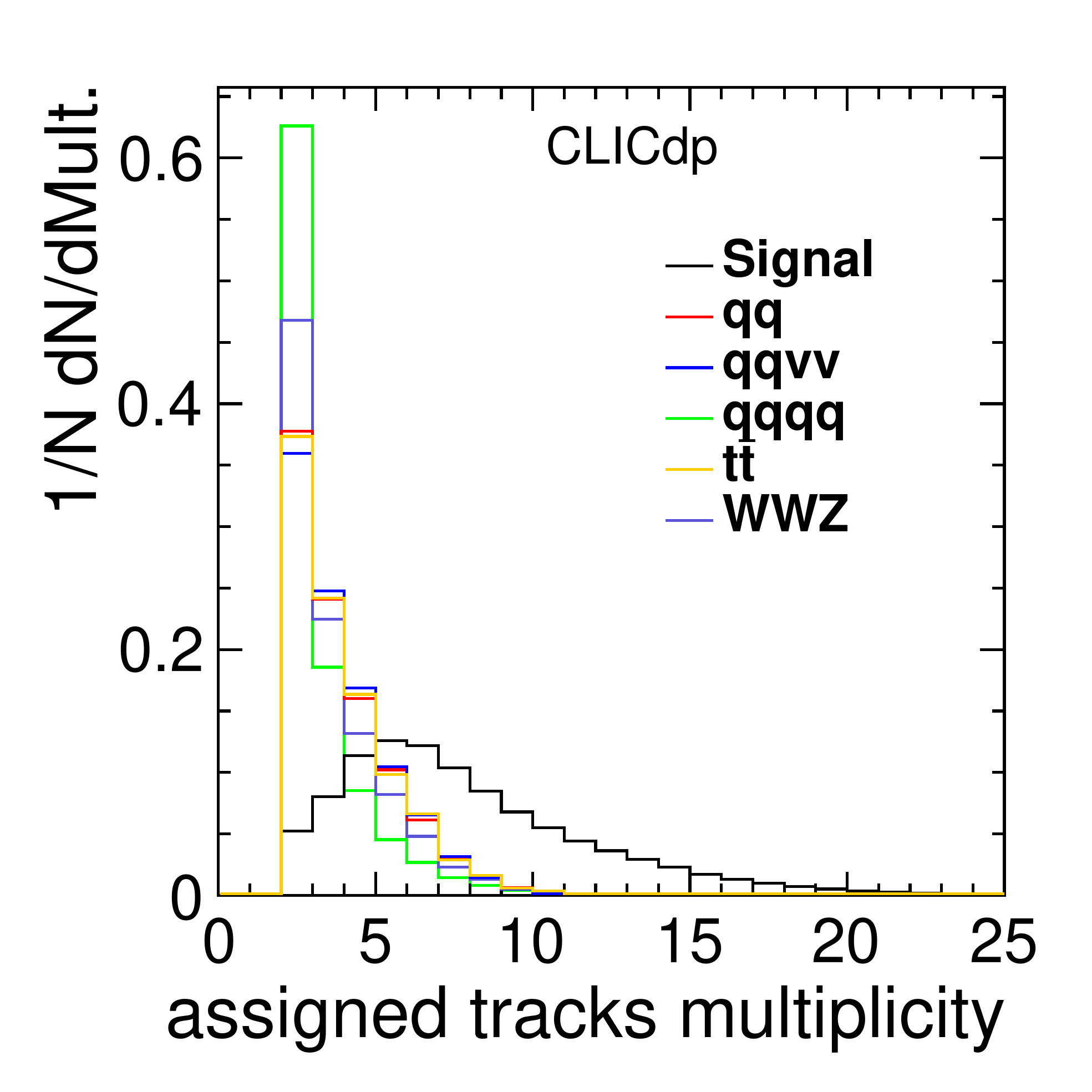}
\put(45,80){\scriptsize$\sqrt{s} = 350$ GeV}%
\end{overpic}
\begin{overpic}[width=0.37\linewidth]{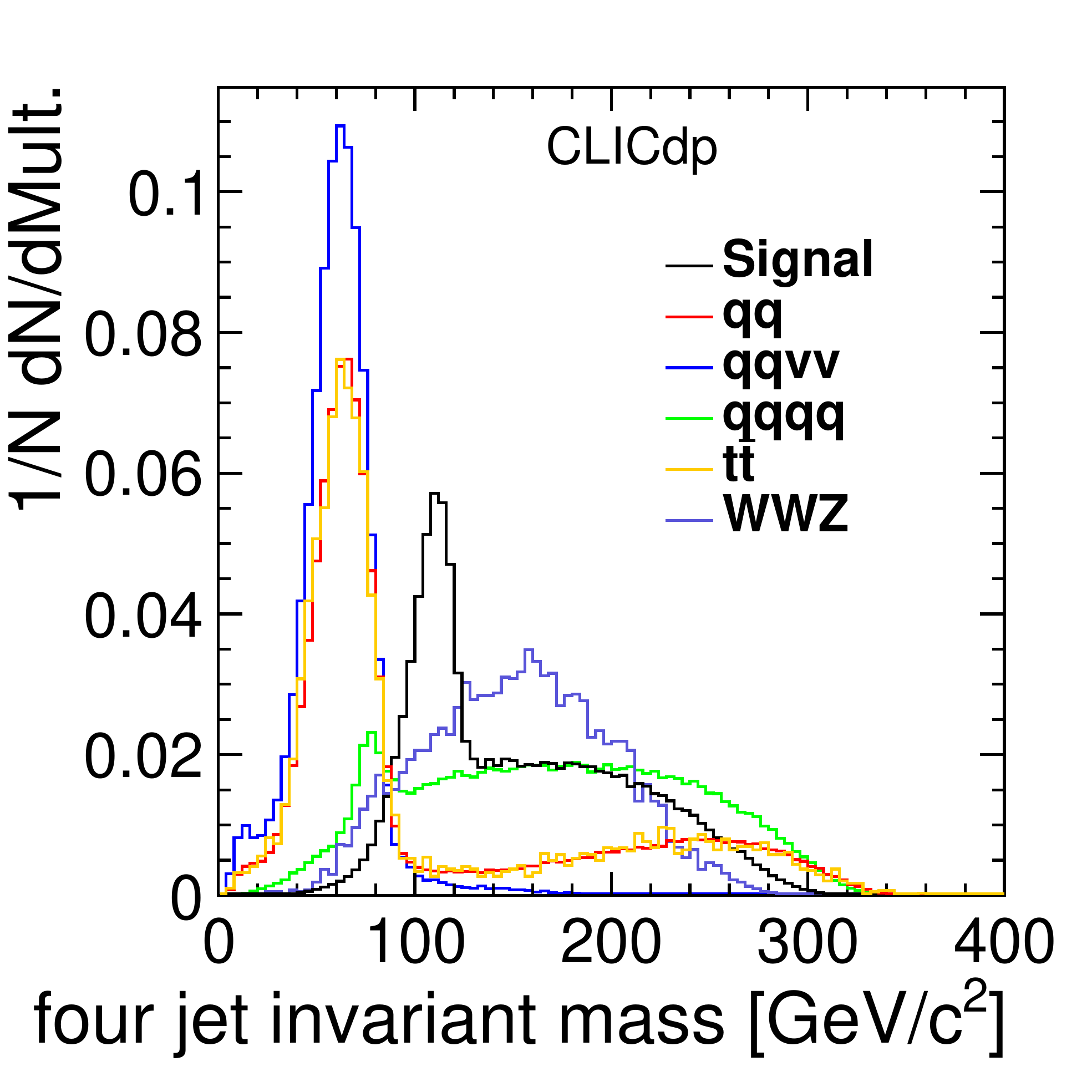}
\put(45,80){\scriptsize$\sqrt{s} = 350$ GeV}%
\end{overpic}
\end{center}
\caption{Number of tracks assigned to the reconstructed DV (left) and four-jet invariant mass (right) for signal samples at $\sqrt{s}=350$~GeV with $\pi^0_v$ mass of 35~GeV/c$^{2}$ and the lifetime of 10~ps, compared to $q \bar{q}$, $q \bar{q}\nu\bar{\nu}$, $q \bar{q} q \bar{q}$, $t \bar{t}$ and $WWZ$ background events.}
\label{fig:BDTGdvmult}
\end{figure}

\begin{figure}[ht]
\begin{center}
\begin{overpic}[width=0.40\linewidth, percent]{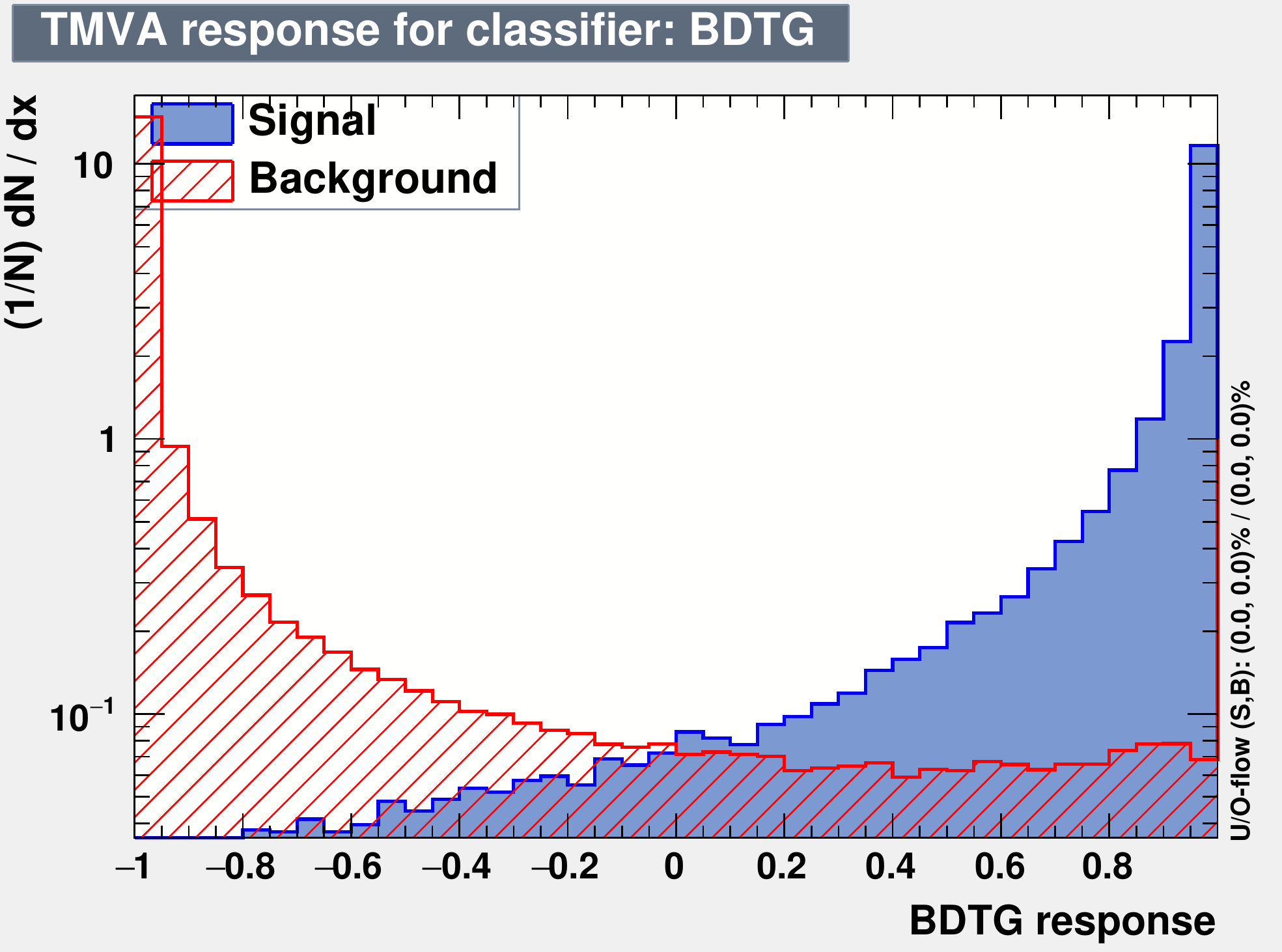}
\put(82,58){(a)}%
\end{overpic}
\begin{overpic}[width=0.40\linewidth, percent]{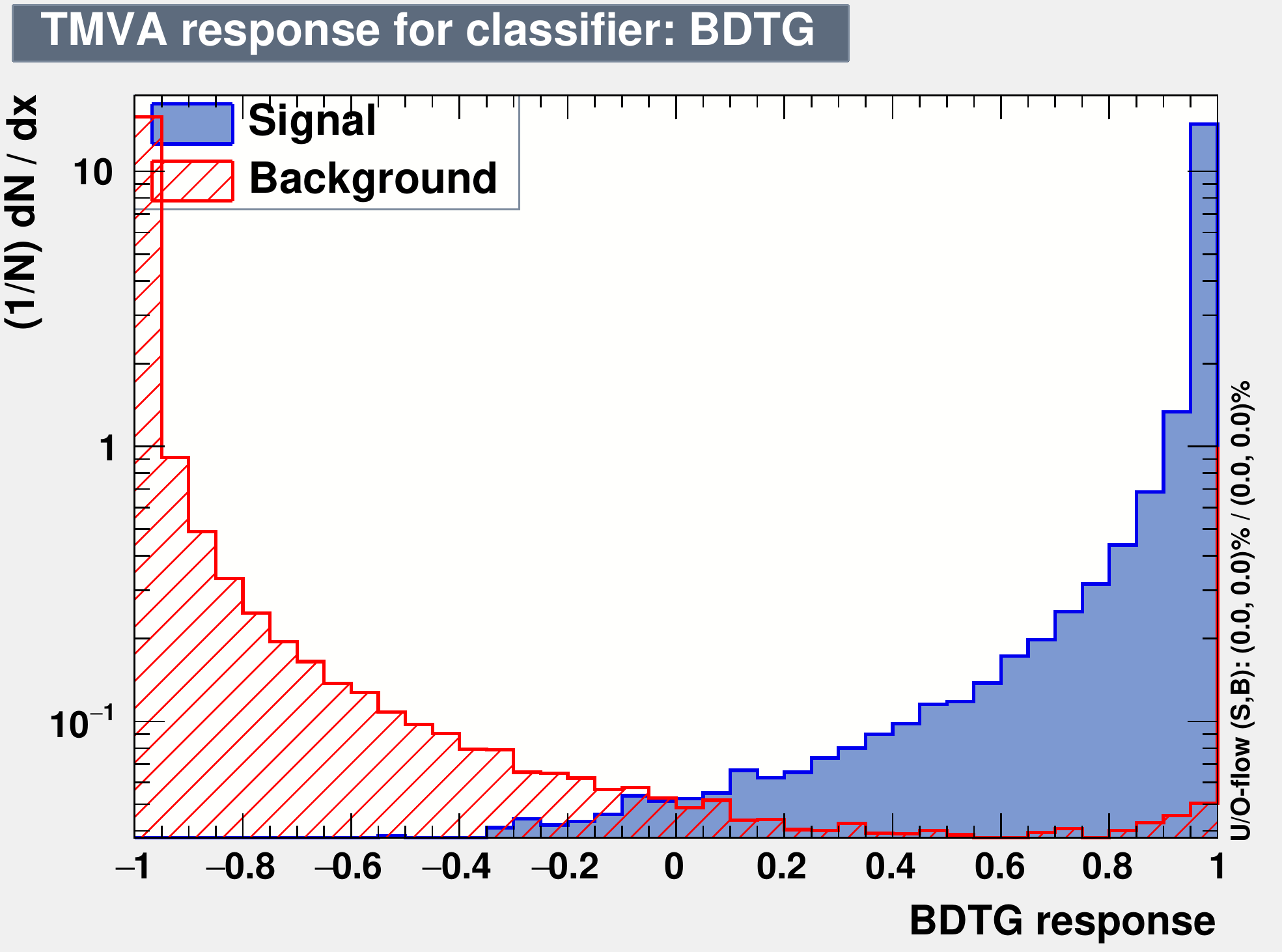}
\put(82,58){(b)}%
\end{overpic}
\begin{overpic}[width=0.40\linewidth, percent]{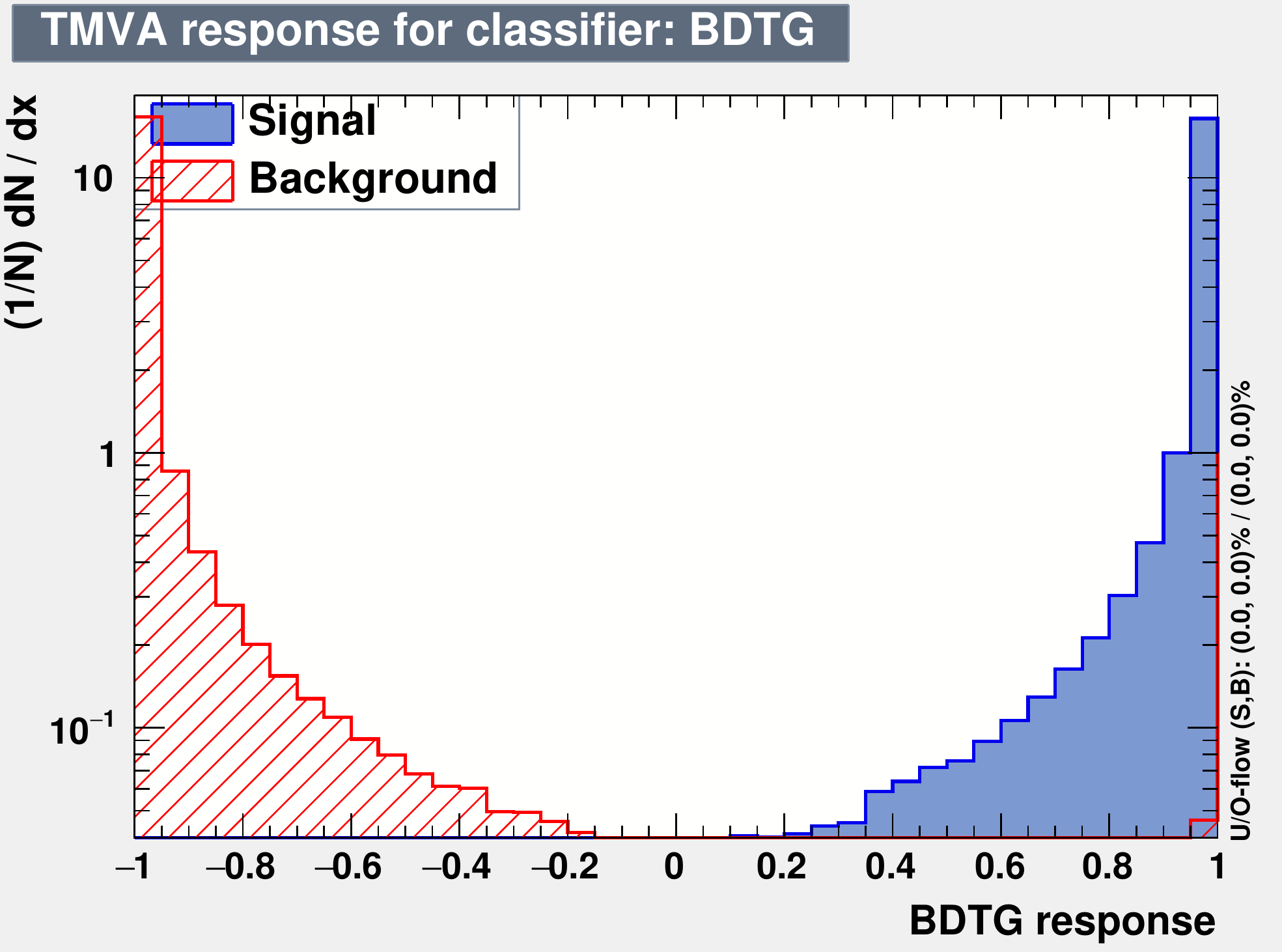}
\put(82,58){(c)}%
\end{overpic}
\begin{overpic}[width=0.40\linewidth, percent]{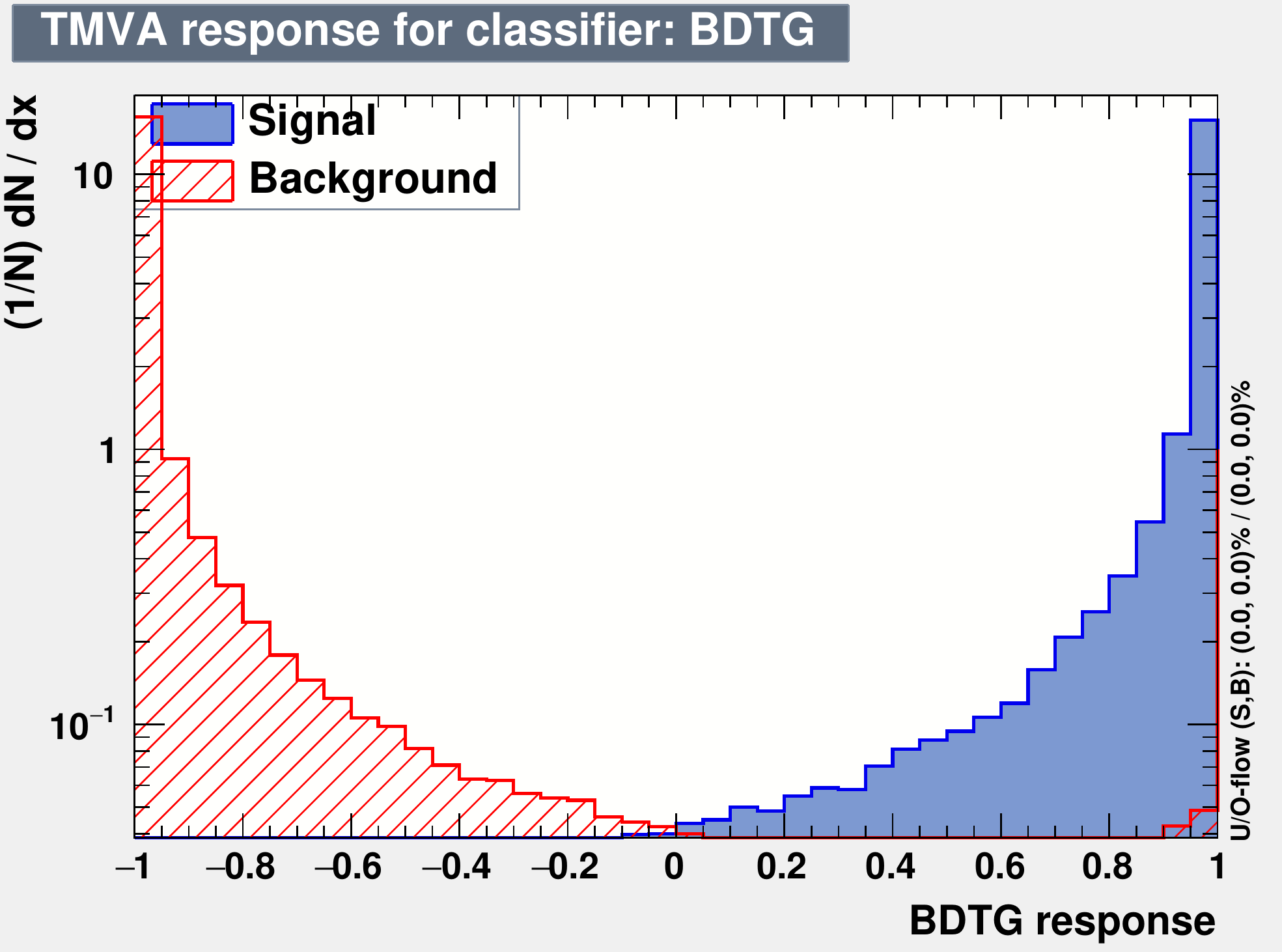}
\put(82,58){(d)}%
\end{overpic}
\end{center}
\caption{Distributions of the BDTG response for signal samples at $\sqrt{s}=350$~GeV with $\pi^0_v$ mass of 35~GeV/c$^{2}$ and four different lifetimes: (a) 1~ps, (b) 10~ps, (c) 100~ps, (d) 300~ps, compared to combined background.}
\label{fig:BDTGresponse}
\end{figure}

As the Hidden Valley objects are predicted to be massive and have non-zero lifetime, the possible presence of new long-lived particles is investigated based on events with reconstructed vertices displaced from the beam axis. In the process $H \to \pi_\textrm{v}^\textrm{0} \pi_\textrm{v}^\textrm{0} \to b \bar{b} b \bar{b}$, in order to reconstruct the parent Higgs boson, two $b$-tagged jets are assigned to each displaced vertex~\cite{CLICHidValley}. The radial distance of the generated $\pi_\textrm{v}^\textrm{0}$ to the beam axis is illustrated in Fig.~\ref{fig:genHVr}, showing the dependence on the $\pi_\textrm{v}^\textrm{0}$ lifetime. The particles are reconstructed using the PANDORA Particle Flow Analysis package~\cite{PFA} with a tight configuration of requirements imposed on the timing and transverse momentum of tracks in order to suppress the beam induced backgrounds. The jets are reconstructed using the longitudinally invariant $k_{t}$ algorithm~\cite{kt} as implemented in the FastJet package~\cite{Fastjet}. The jets are $b$- and $c$-tagged by passing a selection of parameters, including the impact parameters that have been calculated via the vertexing, through a Boosted Decision Tree (BDT)~\cite{BDT}. The $R$ parameter was optimized to minimize the normalized root mean square (RMS/mean) of the di- and four-jet mass. The requirements on the particle impact parameter components as well as the value of $R=1.0$ used for the jet reconstruction are optimized according to the needs of the Hidden Valley analysis (see Ref.~\cite{CLICHidValley} for details). The particles were combined into exactly 4 or 6 jets for the 3~TeV and 350~GeV studies, respectively. Secondary vertices are reconstructed using the LCFI+ (Linear Collider Flavour Identification) package~\cite{LCFI}. An additional requirement on the reconstructed track to have at least one hit in the vertex detector is also imposed. The LCFI+ algorithms, designed primarily to precisely measure individual vertices of $B$ and $D$ hadrons, was found to be inefficient in the reconstruction of displaced vertices from $\pi_\textrm{v}^0$ decays. Therefore, a dedicated procedure to reconstruct displaced vertices has been developed and optimised for the Hidden Valley analysis~\cite{CLICHidValley}. All the jets  are required to have a $b$-tag probability of more than 0.95. Di-jets are constructed by pairing two jets with an invariant mass closest to the $\pi_\textrm{v}^\textrm{0}$ mass. As the jet reconstruction algorithm does not provide the vertex position, a di-jet is assigned to the reconstructed displaced vertex with the largest number of common charged tracks. The efficiencies for the signal and background samples with respect to the requirement to have at least two reconstructed displaced vertices in the event are listed in Table~\ref{tab:samples}. In order to separate the signal from background a multivariate analysis based on the Boosted Decision Tree Gradient (BDTG)~\cite{TMVA} is applied. The BDTG method has been chosen as the most effective, where every signal sample is considered separately and the training is performed for the combined background. The procedure employs seven variables providing significant signal to background separation, i.e. $(i)$ number of tracks assigned to the reconstructed DV, $(ii)$ number of reconstructed DVs in the event, $(iii)$  invariant mass of the DV, $(iv)$ di-jet invariant mass of two jets assigned to the DV, $(v)$ four-jet invariant mass of two di-jets assigned to two DVs, $(vi)$ distance $\textrm{y}_{\textrm{n}+1, \textrm{n}}$ at which the transition from a three-jet event to a two-jet event takes place, $(vii)$ distance $\textrm{y}_{\textrm{n}-1, \textrm{n}}$ at which the transition from a four-jet event to a three-jet event takes place. In addition, the procedure for $\sqrt{s}=350$~GeV includes also the invariant mass of a $Z$ boson candidate, reconstructed from $b$-jets not assigned to any displaced vertex. Figure~\ref{fig:BDTGdvmult} shows  distributions of some variables employed in the BDTG procedure (number of tracks assigned to the reconstructed DV and four-jet invariant mass) for the signal and background samples, being good examples of the separation between the signal and background samples that can be achieved. The BDGT procedure using all the variables listed above allows for a significant signal to background separation, as it can be observed in Fig~\ref{fig:BDTGresponse}, where the distributions of the BDTG response for signal and combined background are shown.

\newpage

\begin{figure}[ht]
\begin{center}
\begin{overpic}[width=0.40\linewidth, percent]{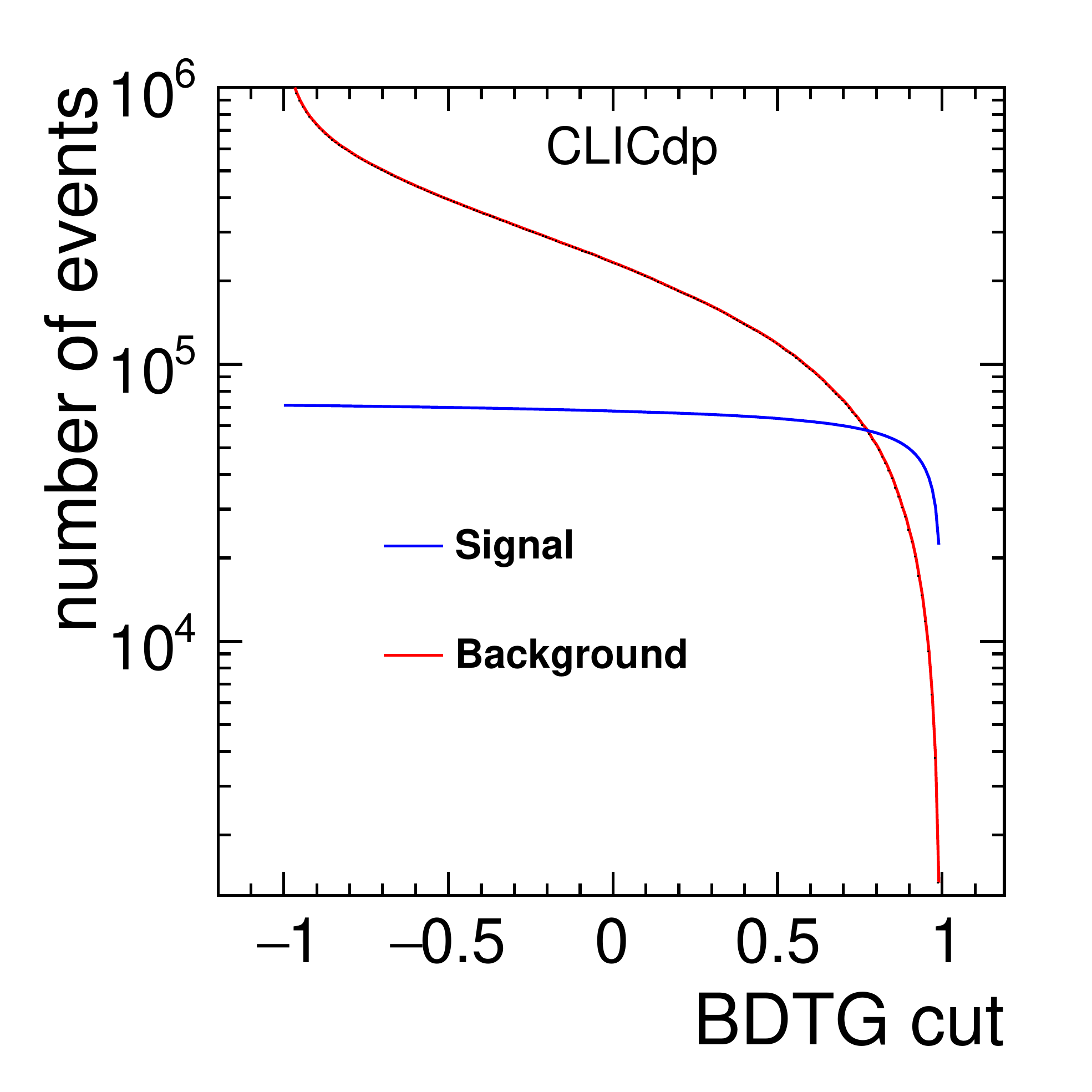}
\put(82,84){(a)}%
\put(45,80){\scriptsize$\sqrt{s} = 350$ GeV}%
\end{overpic}
\begin{overpic}[width=0.40\linewidth, percent]{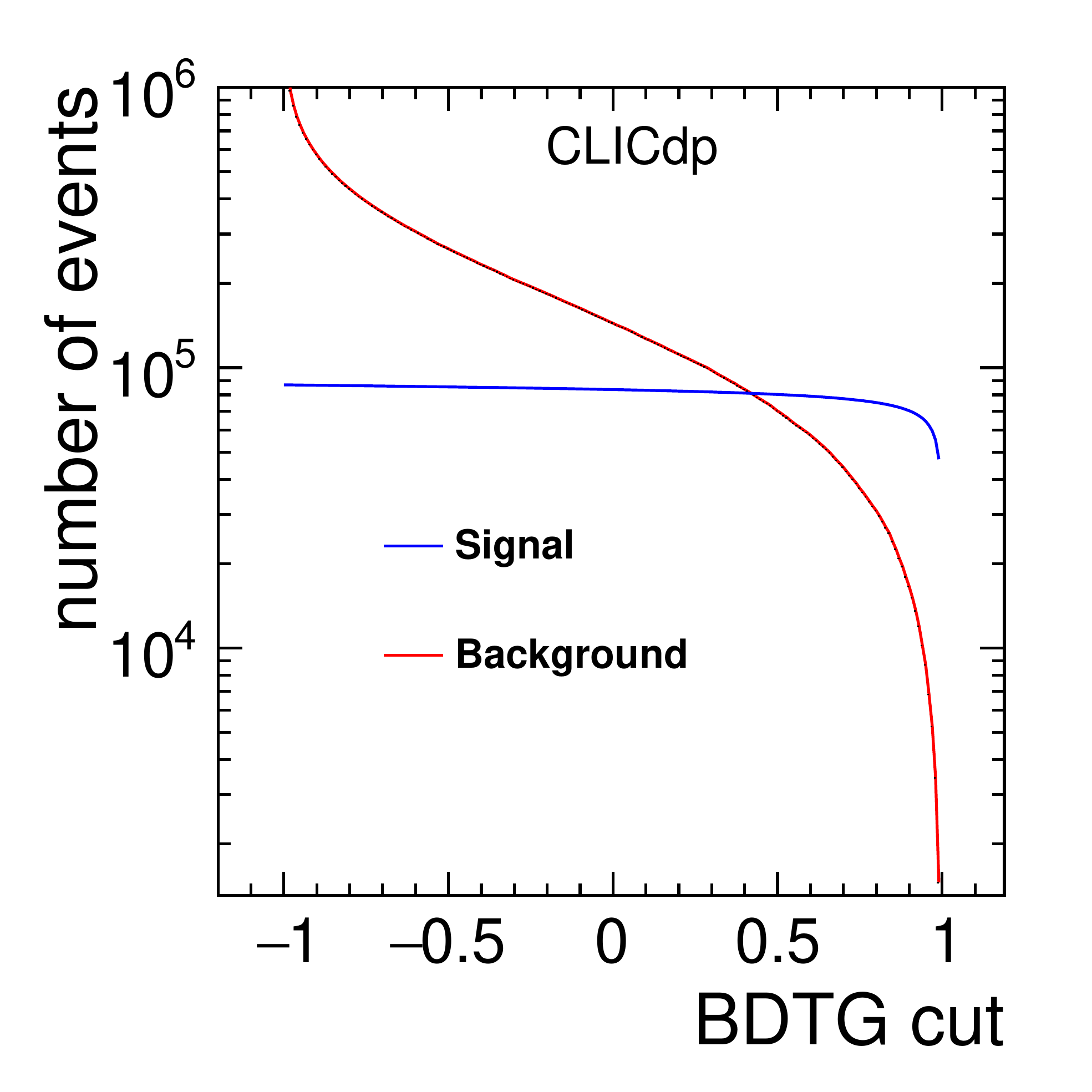}
\put(82,84){(b)}%
\put(45,80){\scriptsize$\sqrt{s} = 350$ GeV}%
\end{overpic}
\begin{overpic}[width=0.40\linewidth, percent]{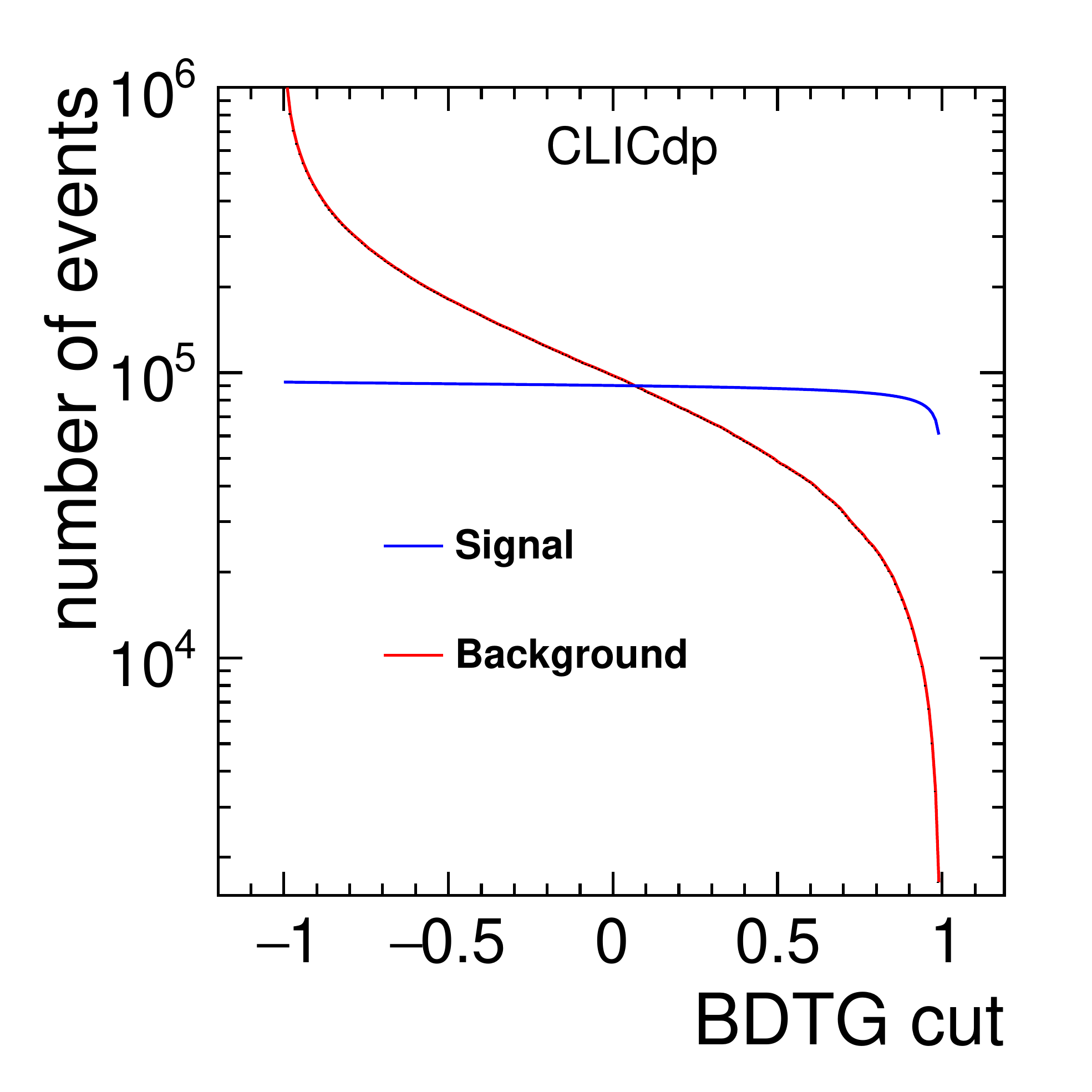}
\put(82,84){(c)}%
\put(45,80){\scriptsize$\sqrt{s} = 350$ GeV}%
\end{overpic}
\begin{overpic}[width=0.40\linewidth, percent]{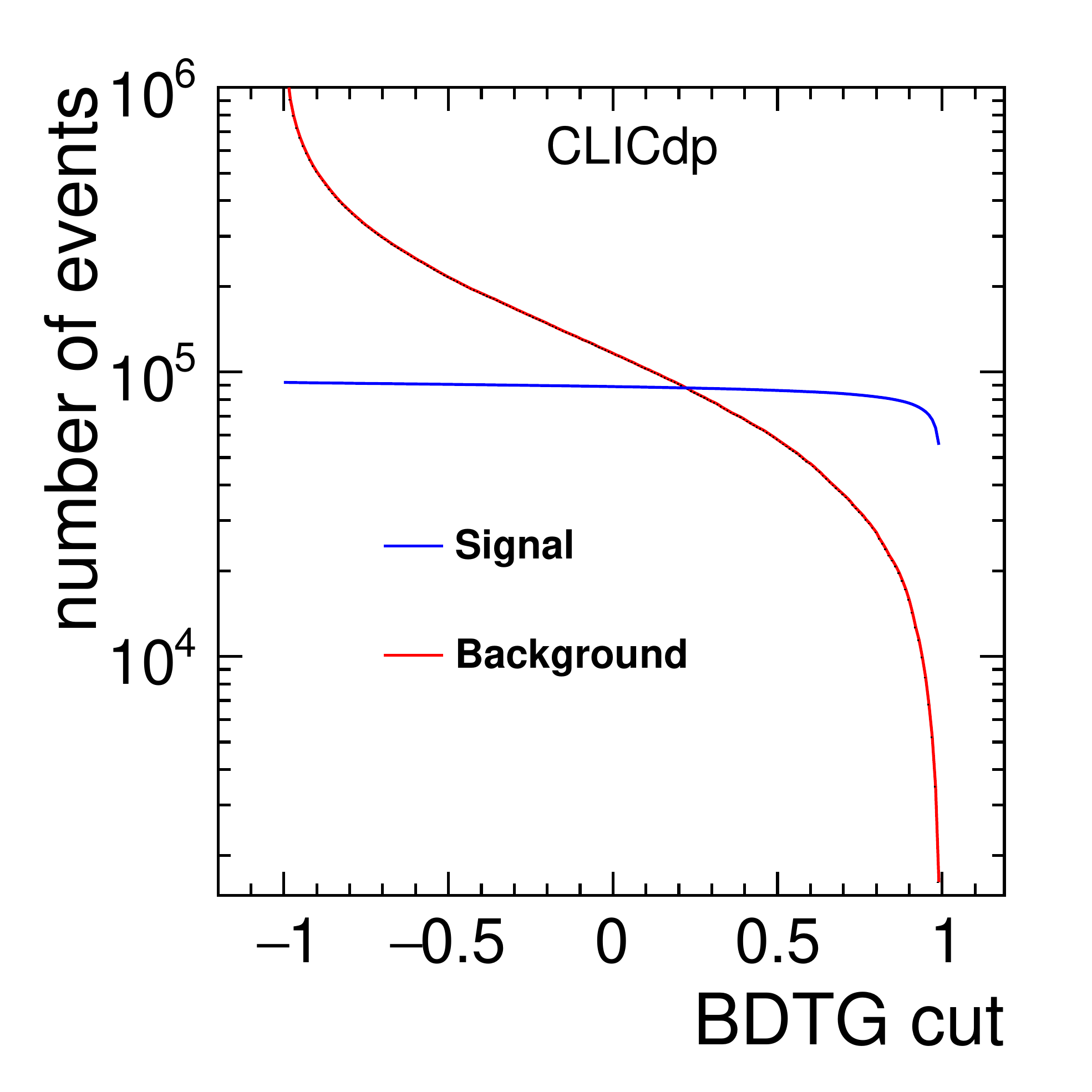}
\put(82,84){(d)}%
\put(45,80){\scriptsize$\sqrt{s} = 350$ GeV}%
\end{overpic}
\end{center}
\caption{Distributions of the number of events left after introducing the requirement on BDTG response for signal samples at $\sqrt{s}=350$~GeV with $\pi^0_v$ mass of 35~GeV/c$^{2}$ and four different lifetimes: (a) 1~ps, (b) 10~ps, (c) 100~ps, (d) 300~ps, compared to combined background.}
\label{fig:BDTGresponseNevt}
\end{figure}
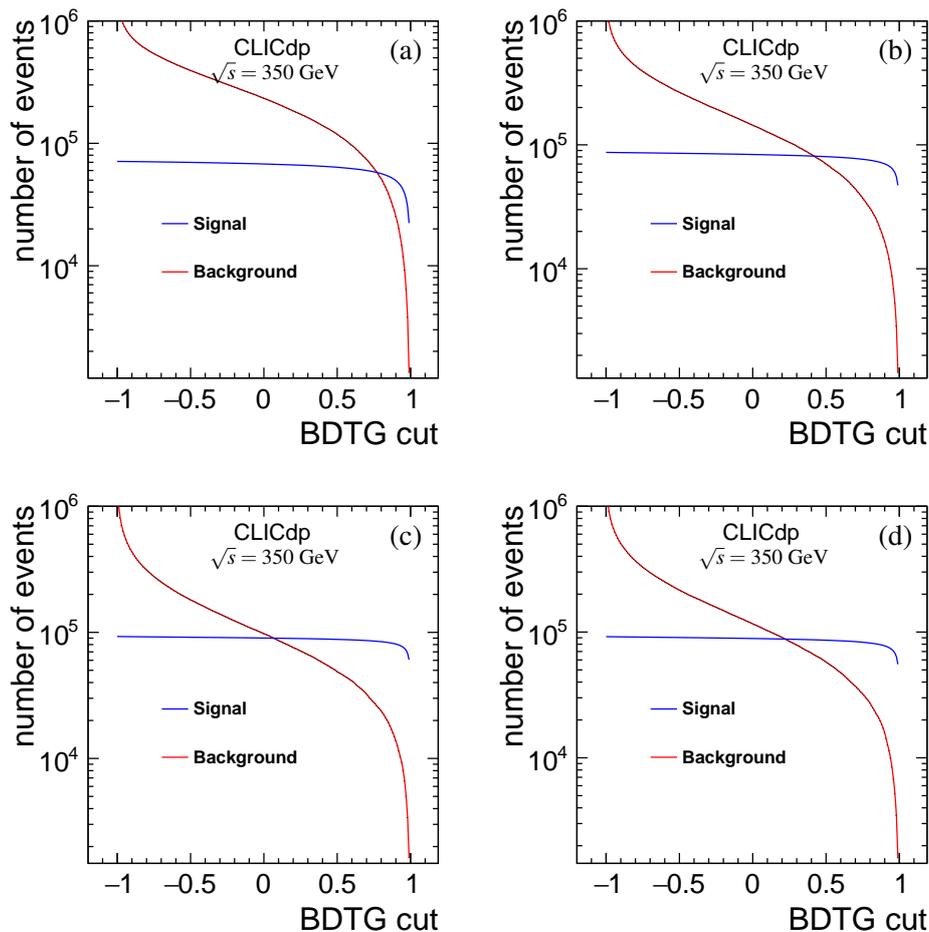

\section{Results}
\begin{figure}[ht]
\begin{center}
\begin{overpic}[width=0.37\linewidth]{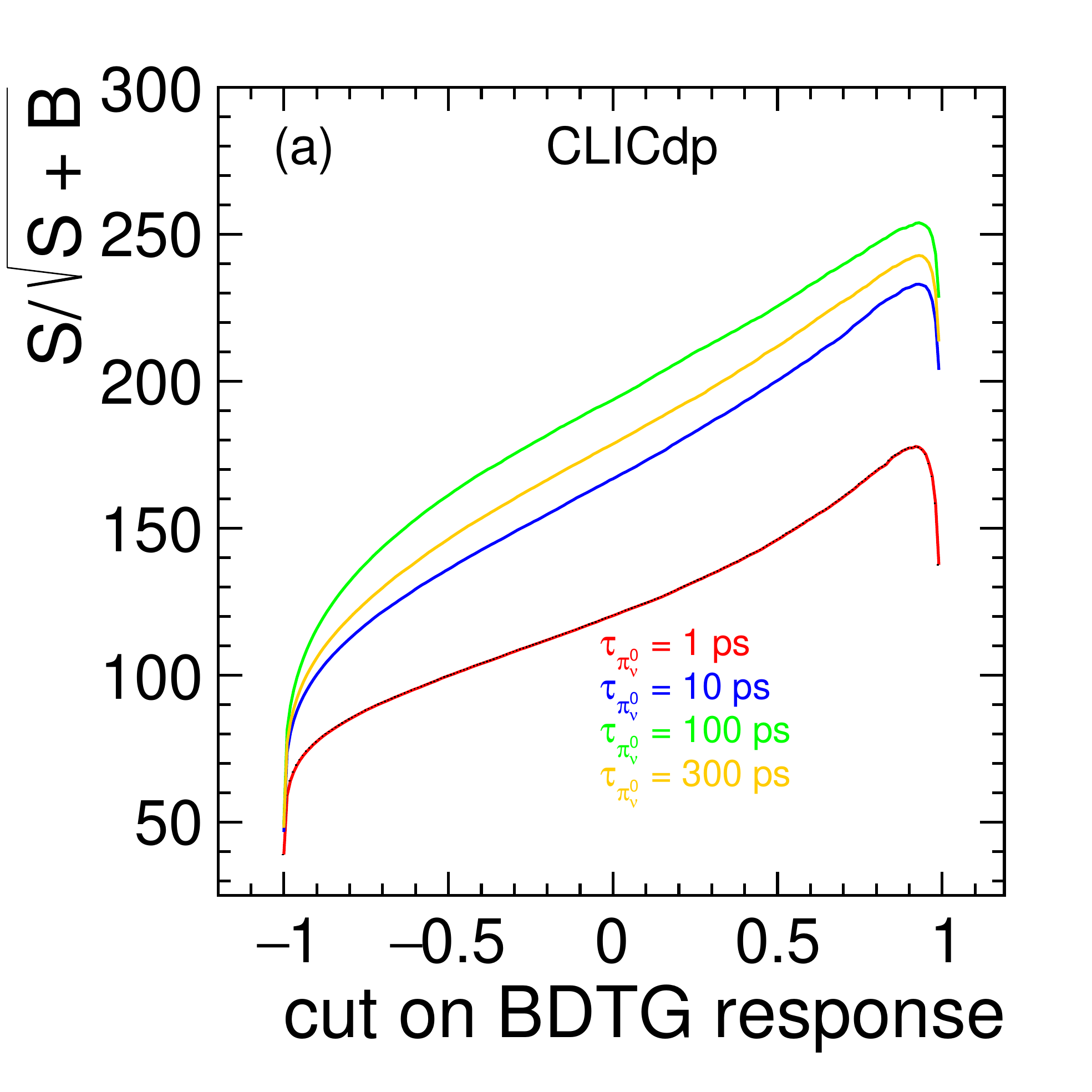}
\put(45,80){\scriptsize$\sqrt{s} = 350$ GeV}%
\end{overpic}
\begin{overpic}[width=0.37\linewidth]{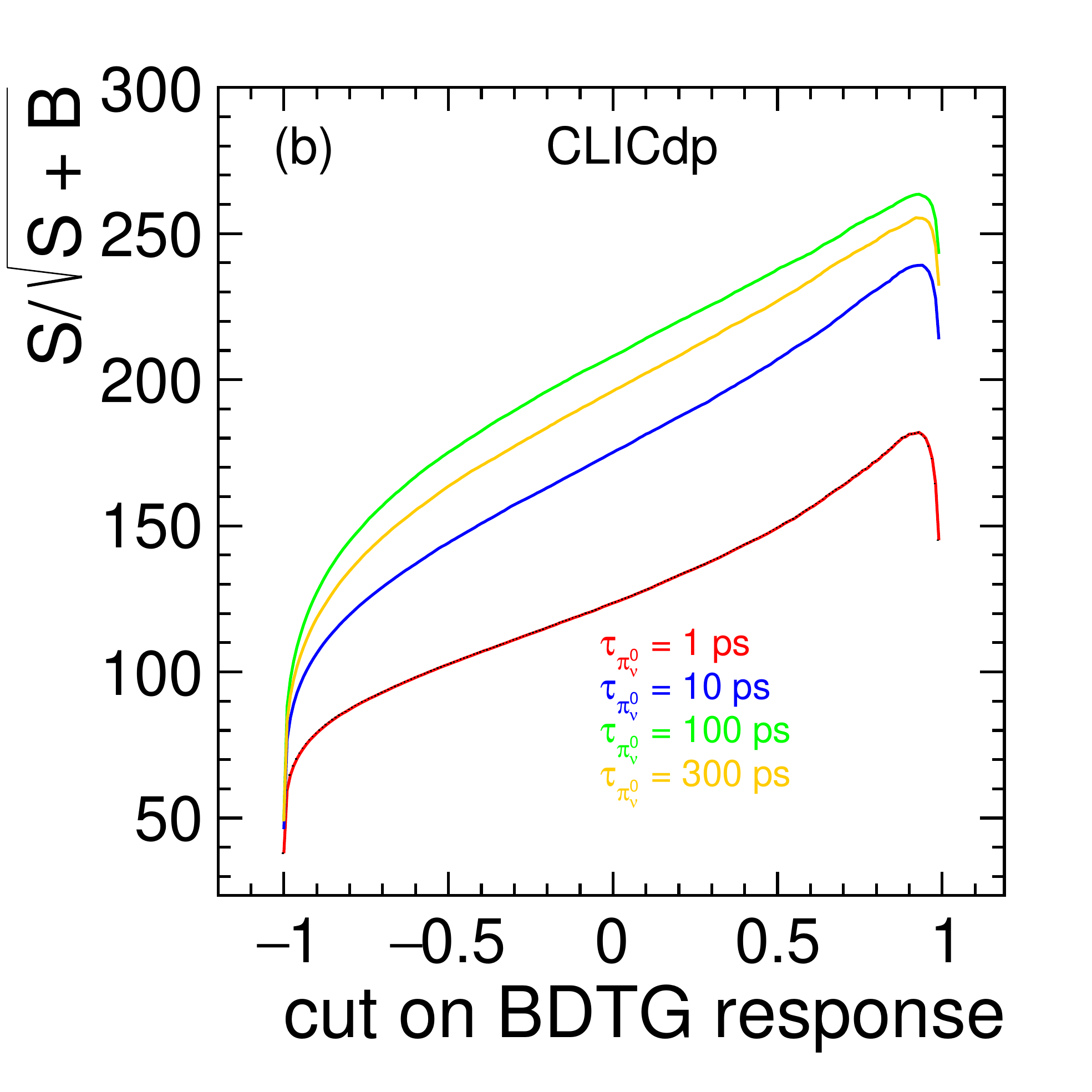}
\put(45,80){\scriptsize$\sqrt{s} = 350$ GeV}%
\end{overpic}
\begin{overpic}[width=0.37\linewidth]{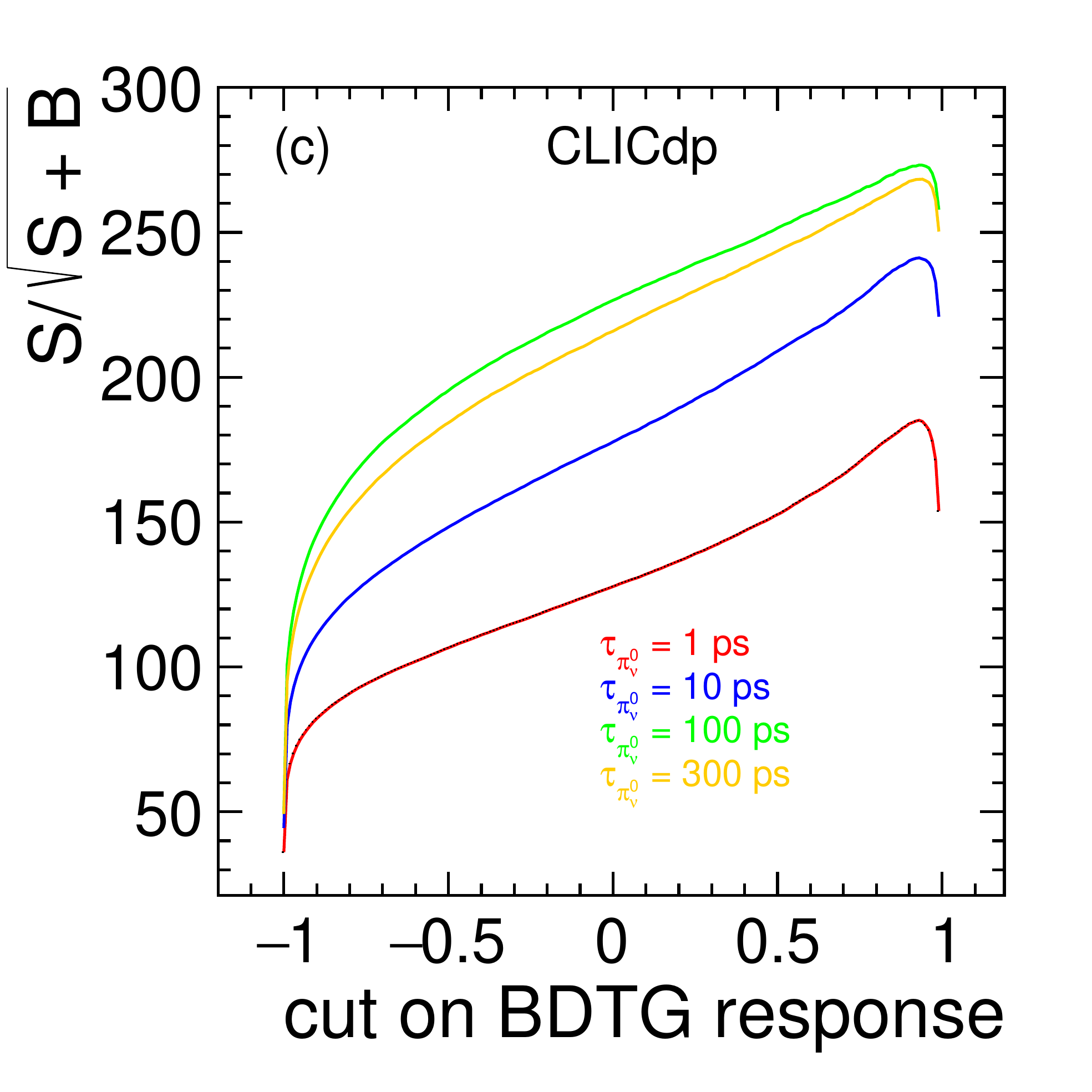}
\put(45,80){\scriptsize$\sqrt{s} = 350$ GeV}%
\end{overpic}
\end{center}
\caption{Sensitivity $S / \sqrt{S+B}$ for the expected number of events as a function of the cut on the BDTG response at $\sqrt{s}=350$~GeV, for signal samples of $\pi^0_v$ with a mass of (a) 25~GeV/c$^{2}$, (b) 35~GeV/c$^{2}$ and (c) 50~GeV/c$^{2}$, and for four different lifetimes: 1~ps (red line), 10~ps (blue line), 100~ps (green line) and 300~ps (yellow line).}
\label{fig:sensitivity1}
\end{figure}

\begin{figure}[ht]
\begin{center}
\begin{overpic}[width=0.37\linewidth]{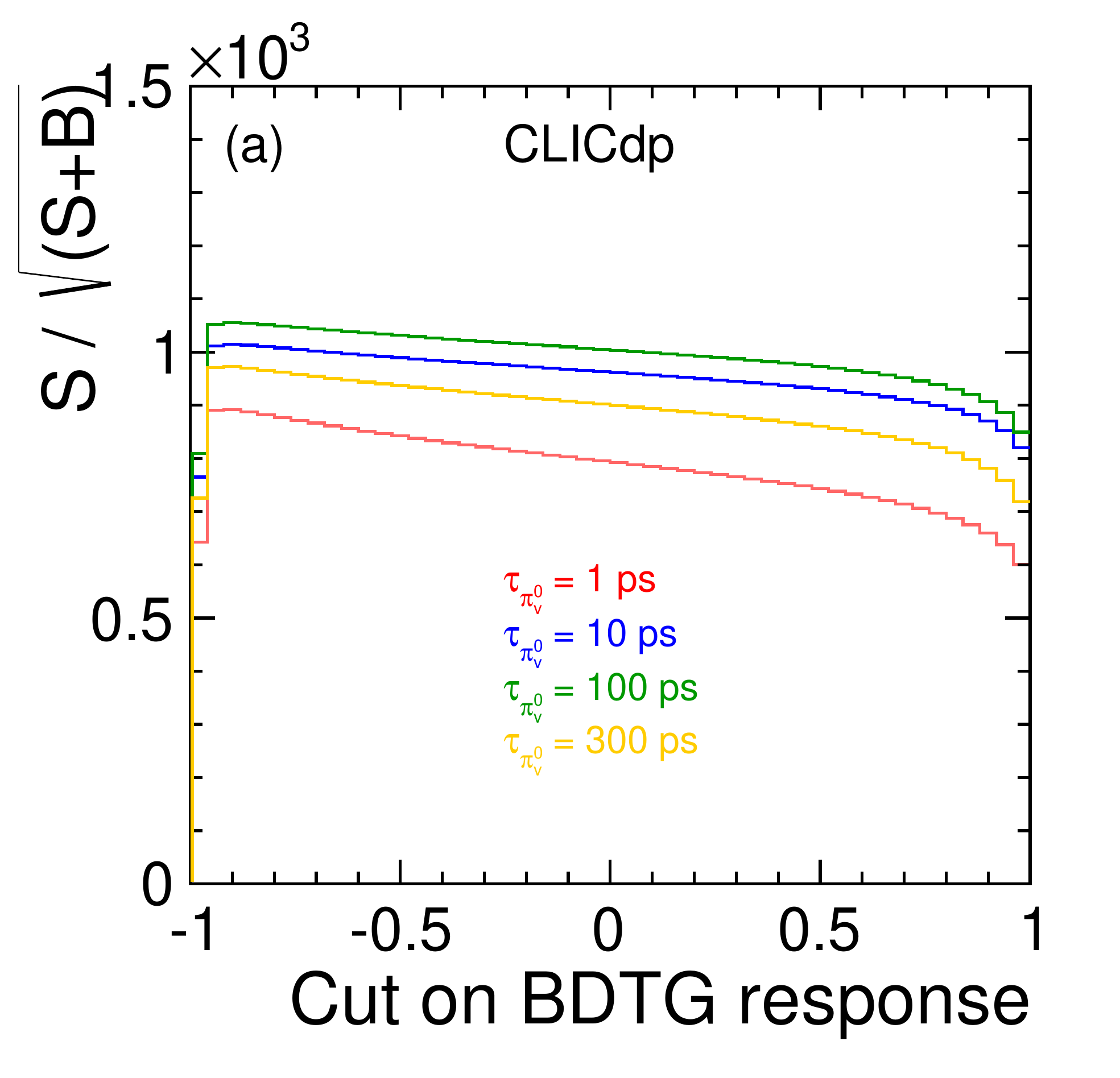}
\put(43,76){\scriptsize$\sqrt{s} = 3$ TeV}%
\end{overpic}
\begin{overpic}[width=0.37\linewidth]{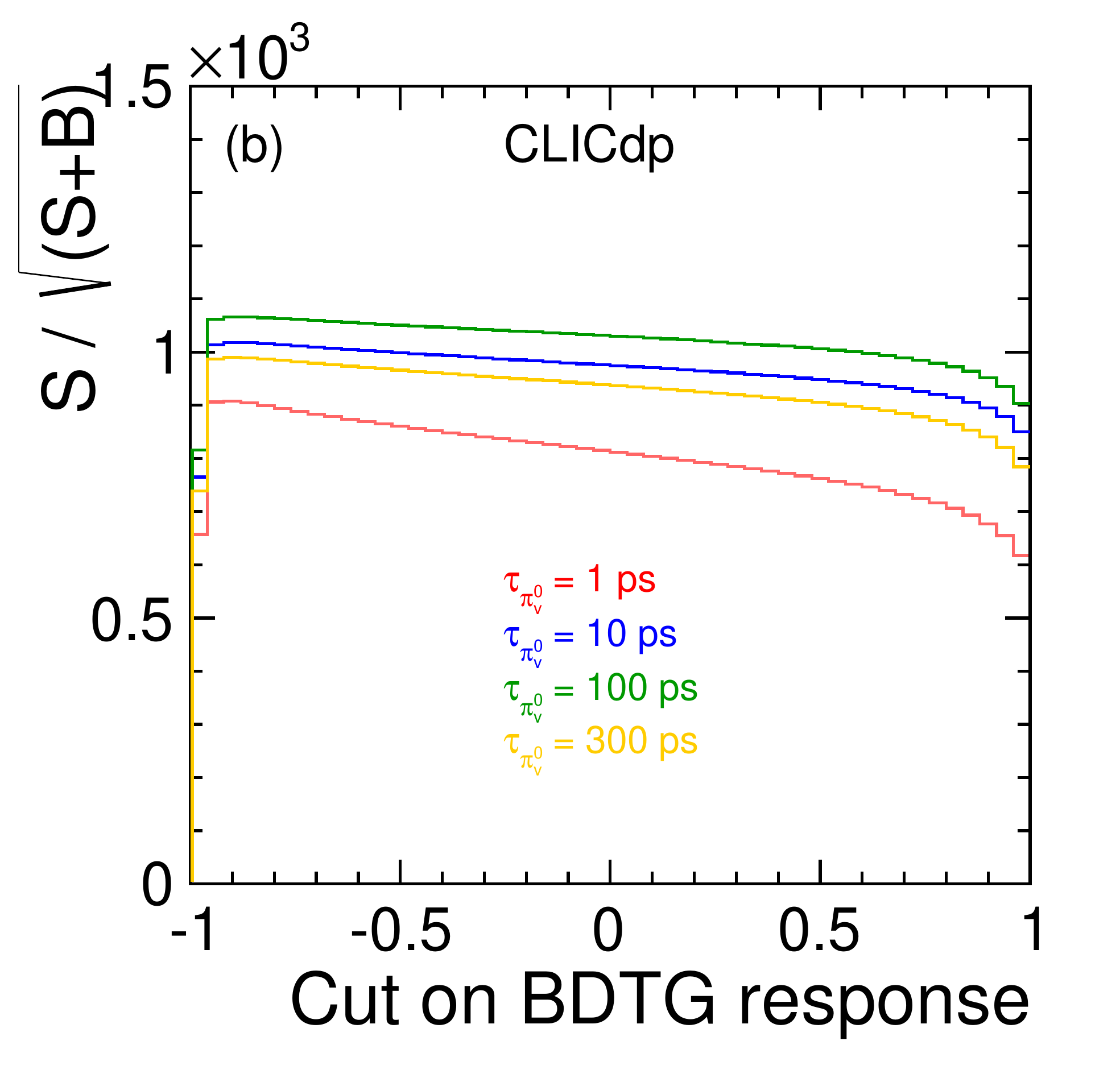}
\put(43,76){\scriptsize$\sqrt{s} = 3$ TeV}%
\end{overpic}
\begin{overpic}[width=0.37\linewidth]{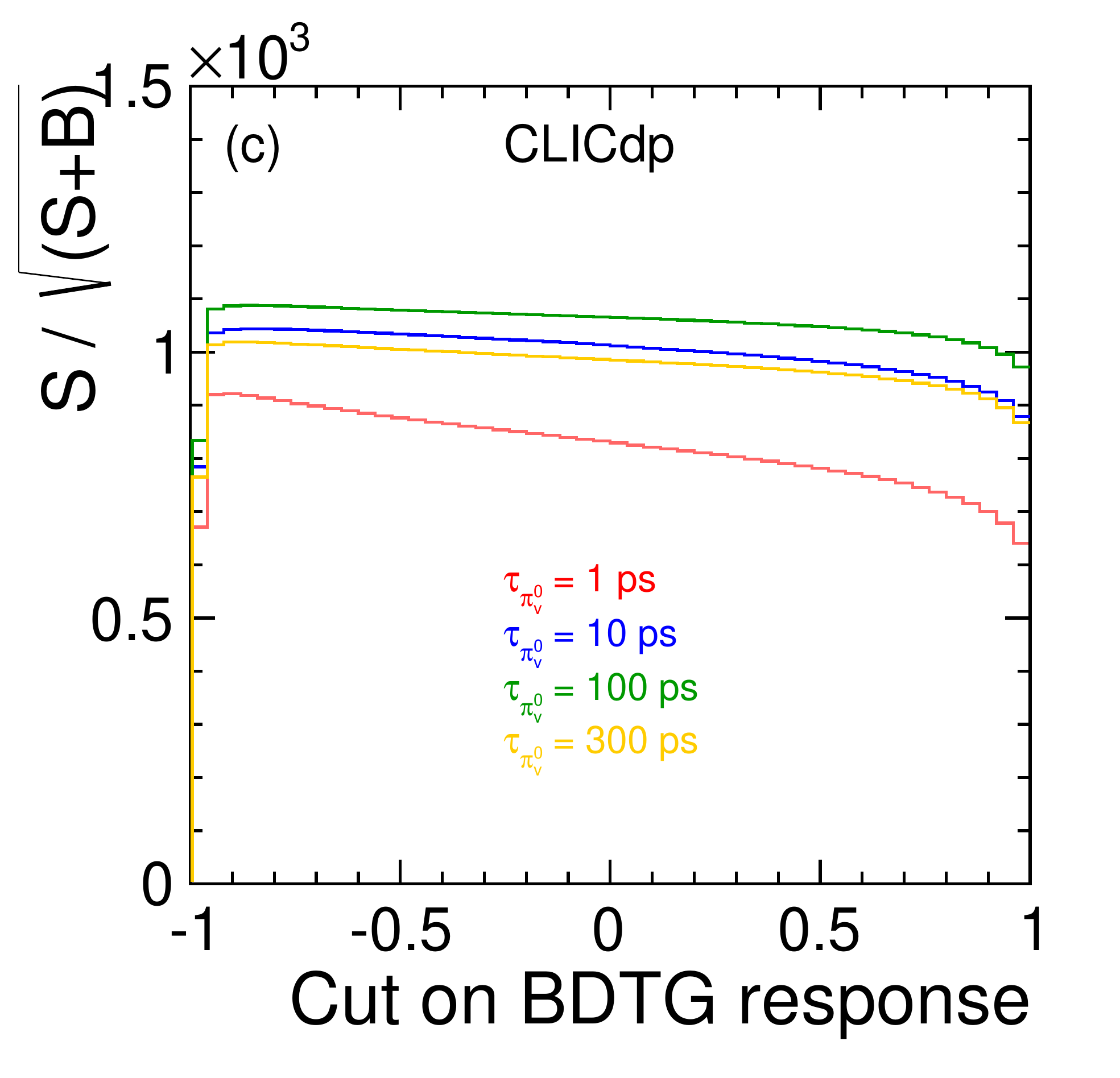}
\put(43,76){\scriptsize$\sqrt{s} = 3$ TeV}%
\end{overpic}
\end{center}
\caption{Sensitivity $S / \sqrt{S+B}$ for the expected number of events as a function of the cut on the BDTG response at $\sqrt{s}=3$~TeV, for signal samples of $\pi^0_v$ with a mass of (a) 25~GeV/c$^{2}$, (b) 35~GeV/c$^{2}$ and (c) 50~GeV/c$^{2}$, and for four different lifetimes: 1~ps (red line), 10~ps (blue line), 100~ps (green line) and 300~ps (yellow line).}
\label{fig:sensitivity2}
\end{figure}

\begin{figure}[ht]
\begin{center}
\begin{overpic}[width=0.37\linewidth]{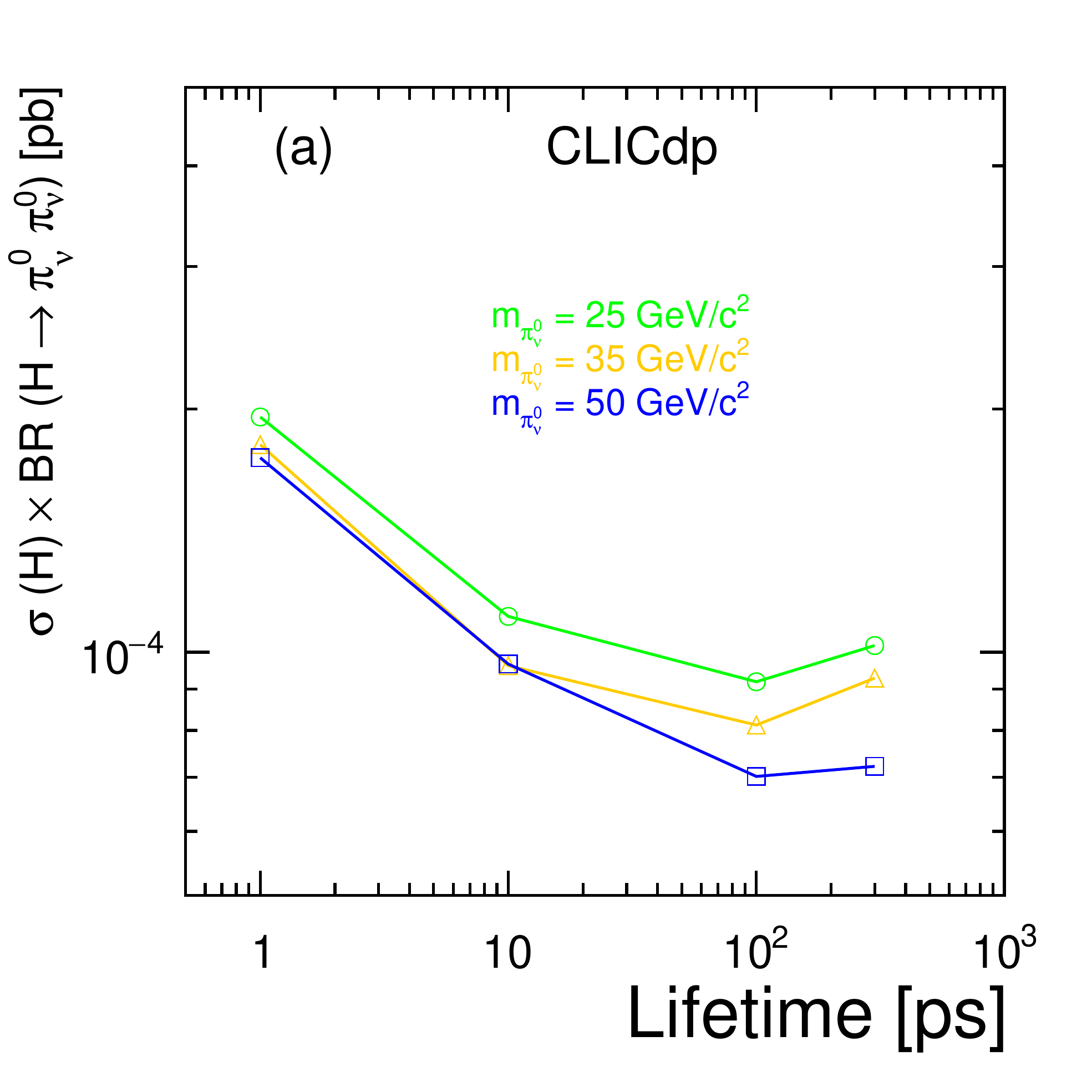}
\put(45,80){\scriptsize$\sqrt{s} = 350$ GeV}%
\end{overpic}
\begin{overpic}[width=0.37\linewidth]{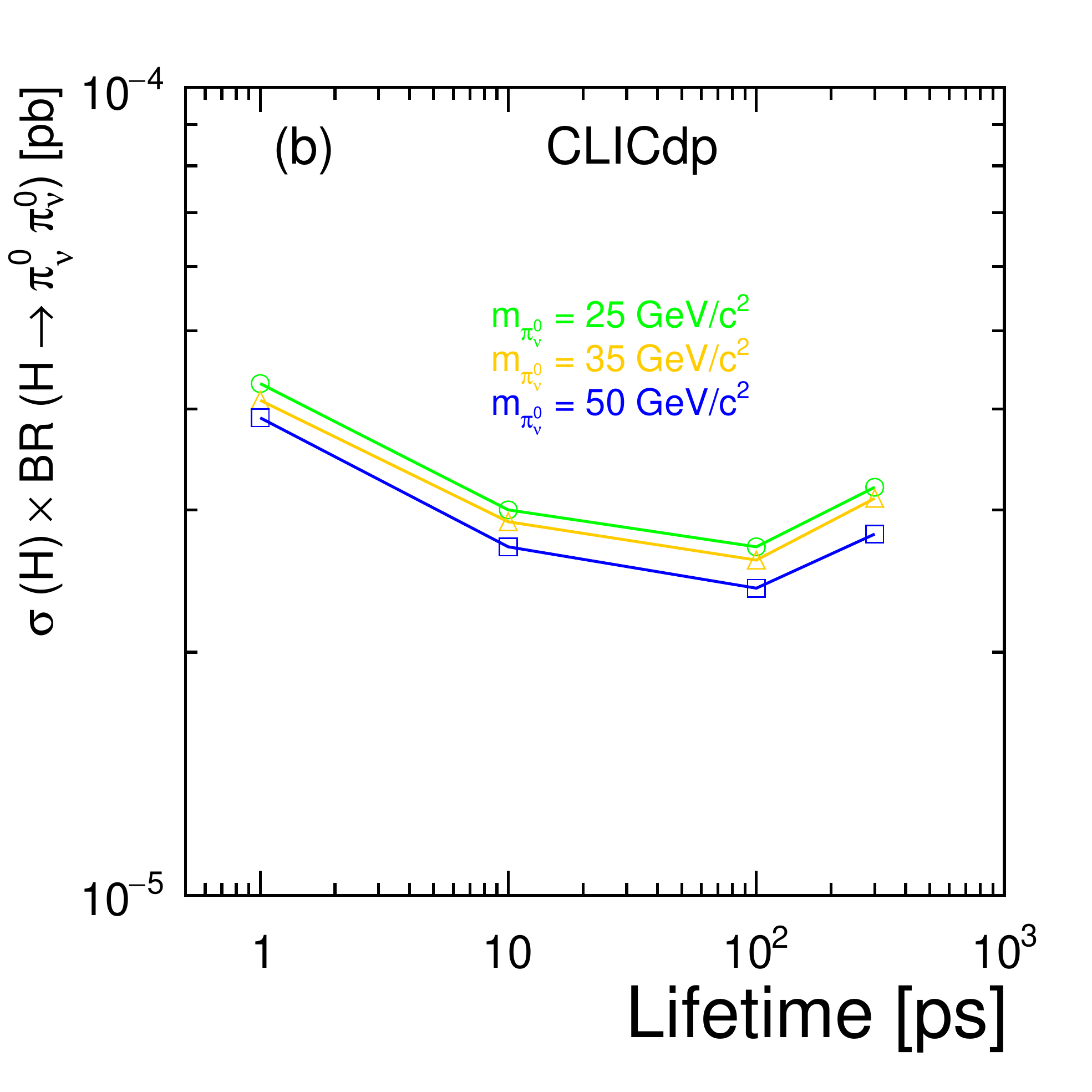}
\put(45,80){\scriptsize$\sqrt{s} = 3$ TeV}%
\end{overpic}
\begin{overpic}[width=0.37\linewidth]{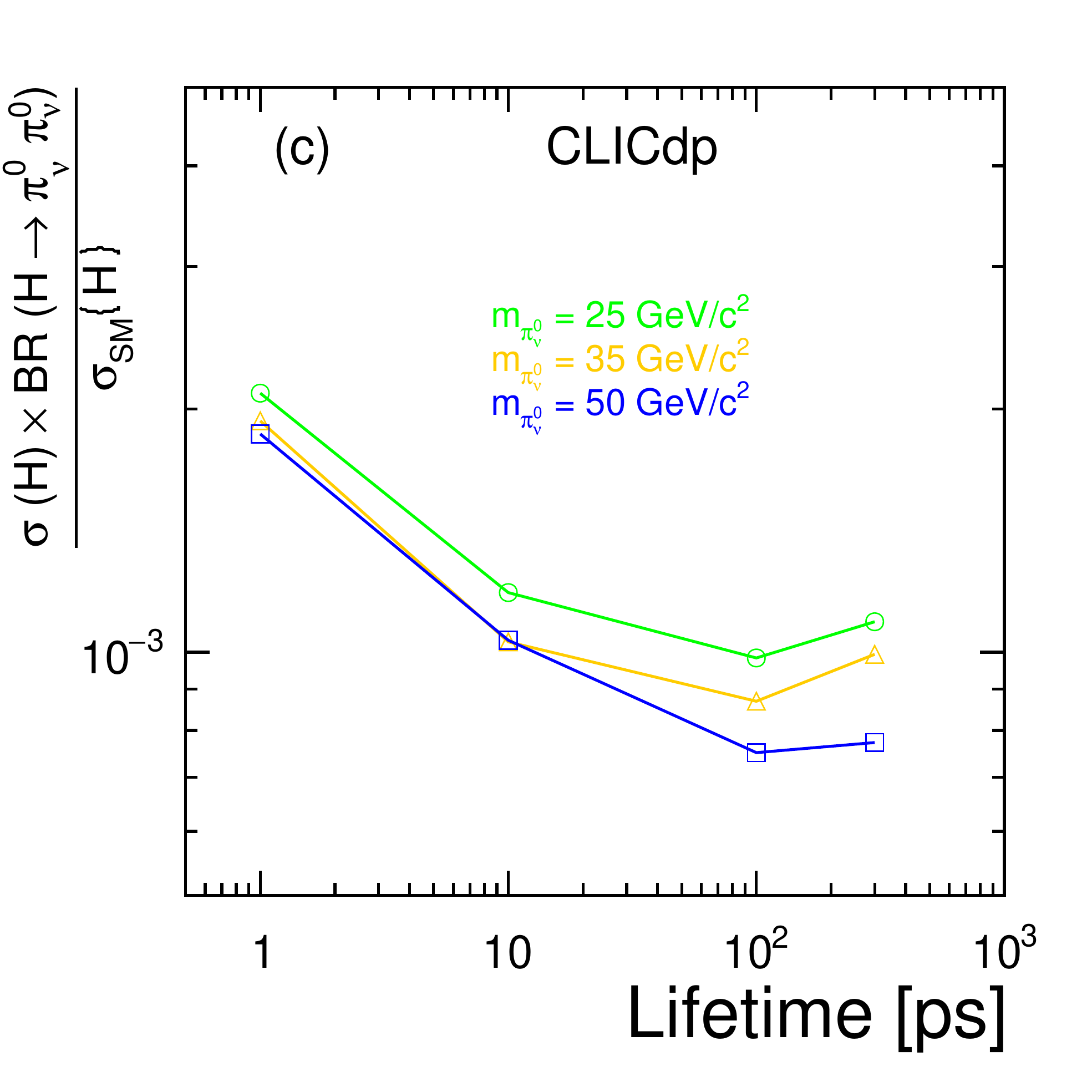}
\put(45,80){\scriptsize$\sqrt{s} = 350$ GeV}%
\end{overpic}
\begin{overpic}[width=0.37\linewidth]{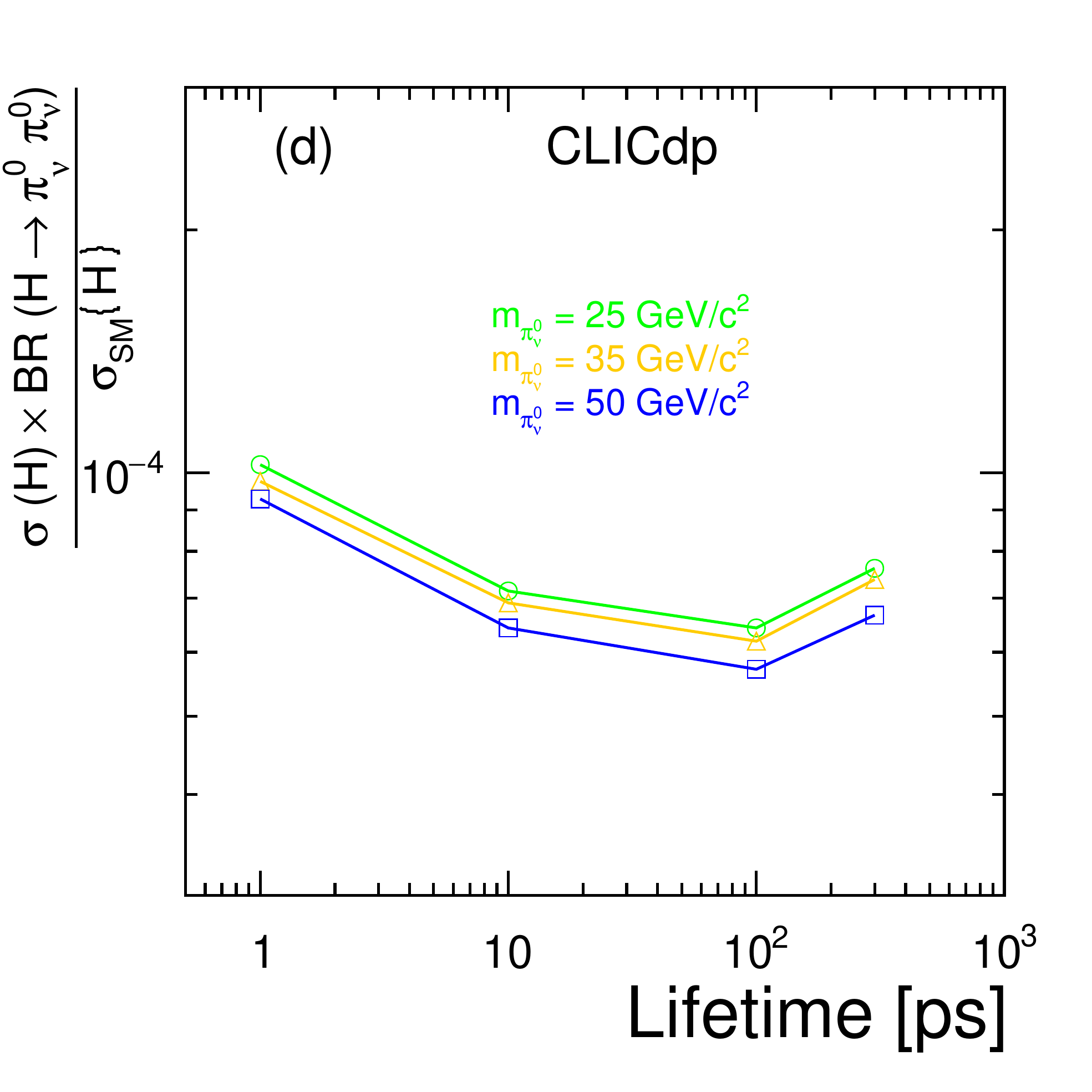}
\put(45,80){\scriptsize$\sqrt{s} = 3$ TeV}%
\end{overpic}
\end{center}
\caption{Expected 95\% CL cross-section upper limits on the $\sigma(H) \times BR(H \rightarrow \pi^0_v \pi^0_v)$, within the model~\cite{hidValley2}, for three different $\pi^0_v$ masses: 25~GeV/c$^{2}$ (green), 35~GeV/c$^{2}$ (yellow), 50~GeV/c$^{2}$ (blue), as a function of $\pi^0_v$ lifetime for $\sqrt{s}=350$~GeV (a) and $\sqrt{s}=3$~TeV (b). The bottom row shows the upper limits normalized to the Standard Model production cross-section of the Higgs boson at $\sqrt{s}=350$~GeV (c) and $\sqrt{s}=3$~TeV (d).}
\label{fig:upl}
\end{figure}

The selection requirement on the BDTG response is optimized to maximize the significance. Its  value was chosen to be > 0.95 for all the $\pi_\textrm{v}^\textrm{0}$ mass and lifetime configurations. The example distributions of the number of signal and combined background events with respect to the requirement on the BDTG response are plotted in Fig.~\ref{fig:BDTGresponseNevt} for $\sqrt{s}=350$~GeV with $\pi^0_v$ mass of 35~GeV/c$^{2}$ and four different lifetimes. For all the $\pi^0_v$ mass and lifetime hypotheses the sensitivity of the CLIC\_ILD detector to observe $\pi^0_v$ particles through the SM Higgs boson decay $H \rightarrow \pi^0_v \pi^0_v$ has been estimated, separately for $\sqrt{s}=350$~GeV and $\sqrt{s}=3$~TeV collision energy, for an  integrated luminosity of 1~ab$^{-1}$ and 3~ab$^{-1}$, respectively. The Higgsstrahlung process was assumed for $\sqrt{s}=350$~GeV, while for $\sqrt{s}=3$~TeV the Higgs boson production via $WW$-fusion was chosen. Fig.~\ref{fig:sensitivity1} shows the sensitivity at $\sqrt{s}=350$~GeV as a function of the cut on the BDTG response for signal samples with different masses and lifetimes, and for the combined background. Similar distributions at $\sqrt{s}=3$~TeV are shown in Fig.~\ref{fig:sensitivity2}. Finally, the expected  upper limits, in the absence of signal observation, on the product of the Higgs production cross-section and the branching fraction of the Higgs boson decay into long-lived particles, $\sigma(H) \times BR(H \rightarrow \pi^0_v \pi^0_v)$ have been determined employing the CL(s) method~\cite{CLs}. Under the assumption of a 100\% branching fraction for $\pi^0_v \rightarrow b\bar{b}$ and the cut on the BDTG response $> 0$, the 95\% CL upper limits on $\sigma(H) \times BR(H \rightarrow \pi^0_v \pi^0_v)$ have been computed. Fig.~\ref{fig:upl} shows the expected 95\% CL cross-section upper limits on the $\sigma(H) \times BR(H \rightarrow \pi^0_v \pi^0_v)$, within the model~\cite{hidValley2}, for three different masses, as a function of $\pi^0_v$ lifetime, at $\sqrt{s}=350$~GeV and $\sqrt{s}=3$~TeV, together with the upper limits normalized to the Standard Model Higgs boson production cross-section.

\newpage

\section{Conclusions}
The sensitivity of CLIC\_ILD detector model to long-lived particles from Higgs boson decays was studied for both the first ($\sqrt{s}$ = 350~GeV) and the last ($\sqrt{s}$ = 3~TeV) stage of its planned operation, with an integrated luminosity of 1~ab$^{-1}$ and 3~ab$^{-1}$, respectively. The dominant production mechanisms ($\textrm{e}^+\textrm{e}^- \to \textrm{Z}(\to \textrm{q}\bar{\textrm{q}})\textrm{H}$ and $\textrm{e}^+\textrm{e}^- \to \textrm{H} \nu_{e} \bar{\nu_{e}}$, respectively) were assumed. The analysis based on reconstructed displaced vertices provides an immense reduction of the large Standard Model background using a multivariate analysis approach. The expected upper limits obtained in the absence of signal observation are much more precise compared to those of the currently operating detectors~\cite{hvATLAS,hvCMS,hvLHCb}.

\section{Acknowledgements}
First of all, we would like to thank Philipp Roloff for his instant support as well as many encouraging discussions. We would also like to thank the whole CLICdp group for the help in various aspects of the analysis, including as well the software related issues.

\newpage

\printbibliography[title=References]

@article{hidValley1,
      author         = "Strassler, M. J. and Zurek, K. M.",
      title          = "{Echoes of a hidden valley at hadron colliders}",
      journal        = "Phys. Lett.",
      volume         = "B651",
      pages          = "374",
      doi            = "10.1016/j.physletb.2007.06.055",
      year           = "2007",
      eprint         = "hep-ph/0604261",
      archivePrefix  = "arXiv",
      primaryClass   = "hep-ph",
}

@article{hidValley2,
      author         = "Strassler, M. J. and Zurek, K. M.",
      title          = "{Discovering the Higgs through highly-displaced vertices}",
      journal        = "Phys. Lett.",
      volume         = "B661",
      pages          = "263",
      doi            = "10.1016/j.physletb.2008.02.008",
      year           = "2008",
      eprint         = "hep-ph/0605193",
      archivePrefix  = "arXiv",
      primaryClass   = "hep-ph",
}

@article{clicILD,
      author         = "{L.~Linssen, A.~Miyamoto, M.~Stanitzki and H. Weerts (eds.)}",
      title          = "{CLIC Conceptual Design Report: Physics and Detectors at CLIC}",
      location       = "CERN",
      year           = "2012",
      eprint         = "1202.5940",
      archivePrefix  = "arXiv",
      primaryClass   = "physics.ins-det",
      series         = "CERN-2012-003",
}

@article{CLICDet,
      author         = "Arominski, D. and others",
      title          = "A detector for CLIC: main parameters and performance",
      journal        = "CLICdp-Note-2018-005",
      doi            = "10.48550/arXiv.1812.07337",
      year           = "2018",
      eprint         = "1812.07337",
      archivePrefix  = "arXiv",
      primaryClass   = "physics.ins-det",
}

@article{CLICD,
      author         = "Munnich, A. and Sailer, A.",
      title          = "The CLIC\_ILD\_CDR Geometry for the CDR Monte Carlo Mass Production",
      journal        = "LCD-Note-2011-002",
      url            = "http://cds.cern.ch/record/1443543",
      year           = "2011",
}

@article{hvCDF,
      author         = "Aaltonen, T. and others",
      title          = "{Search for heavy metastable particles decaying to jet pairs in $p \bar{p}$ at $\sqrt{s}$ = 1.96 TeV}",
      collaboration  = "CDF Collaboration",
      journal        = "Phys. Rev.",
      volume         = "D85",
      pages          = "012007",
      doi            = "10.1103/PhysRevD.85.012007",
      year           = "2012",
      eprint         = "1109.3136",
      archivePrefix  = "arXiv",
      primaryClass   = "hep-ex",
}

@article{hvD0,
      author         = "Abazov, V. M. and others",
      title          = "{Search for resonant pair production of neutral long-Lived particles decaying to $b \bar{b}$ in $p \bar{p}$ collisions at $\sqrt{s}$ = 1.96 TeV}",
      collaboration  = "D0 Collaboration",
      journal        = "Phys. Rev. Lett.",
      volume         = "103",
      pages          = "071801",
      doi            = "10.1103/PhysRevLett.103.071801",
      year           = "2009",
      eprint         = "0906.1787",
      archivePrefix  = "arXiv",
      primaryClass   = "hep-ex",
}

@article{hvATLAS,
      author         = "Aad, G. and others",
      title          = "{Search for a light Higgs boson decaying to long-lived weakly-interacting particles in proton-proton collisions at $\sqrt{s}$ = 7 TeV with the ATLAS detector}",
      collaboration  = "ATLAS Collaboration",
      journal        = "Phys. Rev. Lett.",
      volume         = "108",
      pages          = "251801",
      doi            = "10.1103/PhysRevLett.108.251801",
      year           = "2012",
      eprint         = "1203.1303",
      archivePrefix  = "arXiv",
      primaryClass   = "hep-ex",
}

@article{hvCMS,
      author         = "Khachatryan, V. and others",
      title          = "{Search for long-lived neutral particles decaying to quark-antiquark pairs in proton-proton collisions at $\sqrt{s}$ = 8 TeV}",
      collaboration  = "CMS Collaboration",
      journal        = "Phys. Rev.",
      volume         = "D91",
      pages          = "012007",
      doi            = "10.1103/PhysRevD.91.012007",
      year           = "2015",
      eprint         = "1411.6530",
      archivePrefix  = "arXiv",
      primaryClass   = "hep-ex",
}

@article{hvLHCb,
      author         = "Aaij, R. and others",
      title          = "{Search for long-lived particles decaying to jet pairs}",
      collaboration  = "LHCb Collaboration",
      journal        = "Eur. Phys. J.",
      volume         = "C75",
      pages          = "152",
      doi            = "10.1140/epjc/s10052-015-3344-6",
      year           = "2015",
      eprint         = "1412.3021",
      archivePrefix  = "arXiv",
      primaryClass   = "hep-ex",
}

@article{clicSiD1,
    editor = "Aihara, H. and others",
    title = "{SiD Letter of Intent}",
    eprint = "0911.0006",
    archivePrefix = "arXiv",
    primaryClass = "physics.ins-det",
    reportNumber = "SLAC-R-989, FERMILAB-LOI-2009-01, FERMILAB-PUB-09-681-E",
    month = "11",
    year = "2009"
}

@article{clicSiDILD,
    author = "Abramowicz, Halina and others",
    editor = "Behnke, Ties and Brau, James E. and Burrows, Philip N. and Fuster, Juan and Peskin, Michael and Stanitzki, Marcel and Sugimoto, Yasuhiro and Yamada, Sakue and Yamamoto, Hitoshi",
    title = "{The International Linear Collider Technical Design Report - Volume 4: Detectors}",
    eprint = "1306.6329",
    archivePrefix = "arXiv",
    primaryClass = "physics.ins-det",
    reportNumber = "ILC-REPORT-2013-040, ANL-HEP-TR-13-20, BNL-100603-2013-IR, IRFU-13-59, CERN-ATS-2013-037, COCKCROFT-13-10, CLNS-13-2085, DESY-13-062, FERMILAB-TM-2554, IHEP-AC-ILC-2013-001, INFN-13-04-LNF, JAI-2013-001, JINR-E9-2013-35, JLAB-R-2013-01, KEK-REPORT-2013-1, KNU-CHEP-ILC-2013-1, LLNL-TR-635539, SLAC-R-1004, ILC-HIGRADE-REPORT-2013-003",
    month = "6",
    year = "2013"
}

@article{clicILD2,
    author = "Abe, Toshinori and others",
    collaboration = "Linear Collider ILD Concept Group -",
    title = "{The International Large Detector: Letter of Intent}",
    eprint = "1006.3396",
    archivePrefix = "arXiv",
    primaryClass = "hep-ex",
    reportNumber = "FERMILAB-LOI-2010-03, FERMILAB-PUB-09-682-E, DESY-09-87, KEK-REPORT-2009-6",
    doi = "10.2172/975166",
    month = "2",
    year = "2010"
}

@article{Whizard,
      author         = "Kilian, W. and Ohl, T. and Reuter, J.",
      title          = "{Simulating Multi-Particle Processes at LHC and ILC}",
      journal        = "Eur. Phys. J.",
      volume         = "C71",
      pages          = "1742",
      doi            = "10.1140/epjc/s10052-011-1742-y",
      year           = "2011",
      eprint         = "0708.4233",
      archivePrefix  = "arXiv",
      primaryClass   = "hep-ex",
}

@article{Pythia,
      author         = "{T.~Sjostrand, S.~Mrenna and P.~Skands}",
      title          = "{PYTHIA 6.4 Physics and manual}",
      journal        = "JHEP",
      volume         = "05",
      pages          = "026",
      doi            = "10.1088/1126-6708/2006/05/026",
      year           = "2006",
      eprint         = "hep-ph/0603175",
      archivePrefix  = "arXiv",
      primaryClass   = "hep-ph",
}

@article{Geant4,
      author         = "Allison, J. and others",
      title          = "{Geant4 developments and applications}",
      collaboration  = "Geant4 Collaboration",
      journal        = "IEEE Trans. Nucl. Sci.",
      volume         = "53",
      pages          = "270",
      doi            = "10.1109/TNS.2006.869826",
      year           = "2006",
}

@article{Mokka,
      author         = "Mora de Freitas, P. and Videau, H.",
      title          = "Detector simulation with MOKKA / GEANT4: Present and future",
      journal        = "LC-TOOL-2003-010",
      url            = "http://inspirehep.net/record/609687",
      year           = "2002",
}

@article{Marlin,
      author         = "Gaede, F.",
      title          = "{Marlin and LCCD - Software tools for the ILC}",
      journal        = "Instrum. Meth.",
      volume         = "A559",
      pages          = "177",
      doi            = "10.1016/j.nima.2005.11.138",
      year           = "2006",
}

@article{clicTrack,
      author         = "Gaede, F. and others",
      title          = "{Track reconstruction at the ILC: the ILD tracking software}",
      journal        = "Journal of Physics: Conference Series",
      volume         = "513 2",
      pages          = "022011",
      url            = " http://stacks.iop.org/1742-6596/513/i=2/a=022011",
      year           = "2014",
}

@article{CLICHidValley,
      author         = "Kucharczyk, M. and Wojton, T.",
      title          = "Hidden Valley searches at CLIC",
      journal        = "CLICdp-Note-2018-001",
      url            = "http://cds.cern.ch/record/2625054",
      year           = "2018",
}

@article{PFA,
      author         = "{M.~A.~Thomson, J.~S.~Marshall and A.~Munnich}",
      title          = "{Performance of Particle Flow Calorimetry at CLIC}",
      journal        = "Nucl. Instrum. Meth.",
      volume         = "A700",
      pages          = "153",
      doi            = "10.1016/j.nima.2012.10.038",
      year           = "2013",
      eprint         = "1209.4039",
      archivePrefix  = "arXiv",
      primaryClass   = "physics.ins-det",
}

@article{LCFI,
      author         = "Suehara, T. and Tanabe, T.",
      title          = "{LCFIPlus: A Framework for Jet Analysis in Linear Collider Studies}",
      journal        = "Nucl. Instrum. Meth.",
      volume         = "A808",
      pages          = "109",
      doi            = "10.1016/j.nima.2015.11.054",
      year           = "2016",
      eprint         = "1506.08371",
      archivePrefix  = "arXiv",
      primaryClass   = "physics.ins-det",
}

@article{kt,
      author         = "Catani, S. and others",
      title          = "{Longitudinally-invariant $k_{\perp}$-clustering algorithms for hadron-hadron collisions}",
      journal        = "Nucl. Phys.",
      volume         = "B406",
      pages          = "187",
      doi            = "10.1016/0550-3213(93)90166-M",
      year           = "1993",
}

@article{Fastjet,
      author         = "Cacciari, C. and Salam, G. P. and Soyez, G.",
      title          = "{FastJet user manual}",
      journal        = "Eur. Phys. J.",
      volume         = "C72",
      pages          = "1896",
      doi            = "10.1140/epjc/s10052-012-1896-2",
      year           = "2012",
      eprint         = "1111.6097",
      archivePrefix  = "arXiv",
      primaryClass   = "hep-ph",
}

@article{BDT,
      author         = "{J.~Zhu, B.~P.~Roe, H.-J.~Yang, Y.~Liu, I.~Stancu, and G.~McGregor}",
      title          = "{Boosted Decision Trees as an Alternative to Artificial Neural Networks for Particle Identification}",
      journal        = "Nucl. Instrum. Meth.",
      volume         = "A543",
      pages          = "577-584",
      doi            = "10.1016/j.nima.2004.12.018",
      year           = "2005",
      eprint         = "physics/0408124",
      archivePrefix  = "arXiv",
      primaryClass   = "physics.data-an",
}

@article{TMVA,
      author         = "Voss, H. and others",
      title          = "{TMVA, the Toolkit for Multivariate Data Analysis with ROOT}",
      journal        = "PoS ACAT",
      volume         = "184",
      year           = "2007",
      url            = "https://pos.sissa.it/050/040/pdf",
}

@article{CLs,
      author         = "A.~L.~Read",
      title          = "{Presentation of search results: The CL(s) technique}",
      journal        = "J. Phys.",
      volume         = "G28",
      pages          = "2693",
      doi            = "10.1088/0954-3899/28/10/313",
      year           = "2002",
}

@article{hgg,
      author         = "Barklow, T and Dannheim, D and Sahin, M and Schulte, D",
      title          = "Simulation of $\gamma \gamma \rightarrow$~hadrons background at CLIC",
      journal        = "LCD-Note-2011-020",
      url            = "https://cds.cern.ch/record/1443518",
      year           = "2012",
}

\end{document}